\documentclass[trackchanges]{aastex701}

\usepackage{booktabs}
\usepackage{tabularx}
\usepackage{enumitem}
\usepackage{multirow}
\usepackage{commath}
\usepackage{xcolor}
\usepackage{amsmath}
\usepackage{graphicx}
\usepackage{float}
\usepackage{rotating}
\usepackage{comment}
\usepackage{tikz}
\usetikzlibrary{positioning,arrows}
\tikzset{>=stealth', line cap=round}
\tikzstyle{block}   = [rectangle, rounded corners, draw, fill=orange!20,
                        minimum width=3.0cm, minimum height=1.0cm, align=center, font=\scriptsize]
\tikzstyle{tall}    = [rectangle, rounded corners, draw, fill=orange!20,
                        minimum width=3.2cm, minimum height=3.0cm, align=center, font=\scriptsize]
\tikzstyle{output}  = [rectangle, rounded corners, draw, fill=red!20,
                        minimum width=3.2cm, minimum height=1.1cm, align=center, font=\scriptsize]
\tikzstyle{arrow}   = [thick,->]

\newcolumntype{C}{>{\centering\arraybackslash}X}

\begin{document}

\title{Looking at infrared background radiation anisotropies with Spitzer II. Small scale anisotropies and their implications for new and upcoming space surveys.}

\author[0009-0000-3672-0198]{Aidan J. Kaminsky}
\email{ajk324@miami.edu}
\affiliation{Department of Physics, University of Miami, Coral Gables, FL 33124, USA}
\affiliation{Code 665, Observational Cosmology Lab, NASA Goddard Space Flight Center, Greenbelt, MD 20771, USA}
\author[0000-0003-2156-078X]{Alexander Kashlinsky}
\email{alexander.kashlinsky-1@nasa.gov}
\affiliation{Dept of Astronomy, University of Maryland, College Park, MD 20742, USA}
\affiliation{Code 665, Observational Cosmology Lab, NASA Goddard Space Flight Center, Greenbelt, MD 20771, USA}
\affiliation{Center for Research and Exploration in Space Science and Technology, NASA/GSFC, Greenbelt, MD 20771, USA}
\author[0000-0001-8403-8548]{Richard G. Arendt}
\email{Richard.G.Arendt@nasa.gov}
\affiliation{Center for Research and Exploration in Space Science and Technology, NASA/GSFC, Greenbelt, MD 20771, USA}
\affiliation{Code 665, Observational Cosmology Lab, NASA Goddard Space Flight Center, Greenbelt, MD 20771, USA}
\affiliation{Center for Space Sciences and Technology, University of Maryland, Baltimore County, Baltimore, MD 21250, USA}
\author[0000-0002-1697-186X]{Nico Cappelluti}
\email{ncappelluti@miami.edu}
\affiliation{Department of Physics, University of Miami, Coral Gables, FL 33124, USA}

\begin{abstract}
Spitzer-based source-subtracted cosmic infrared background (CIB) fluctuations at arcminute-to-degree scales indicate the presence of new populations, whereas sub-arcminute power arises from known $z\lesssim 6$ galaxies. We reconstruct the evolution of the near-IR CIB anisotropies on sub-arcminute scales by known galaxy populations. This method is based on, and significantly advanced over, the empirical reconstruction by  \cite{Helgason2012} which is combined with the halo model connecting galaxies to their host dark matter (DM) halos. The modeled CIB fluctuations from known galaxies produce the majority of the observed small-scale signal down to statistical uncertainties of $< 10\%$ and we constrain the evolution of the halo mass regime hosting such galaxies. Thus the large-scale CIB fluctuations from new populations are produced by sources with negligible small-scale power. This appears to conflict with the presented Intra-halo light (IHL) models, but is accounted for if the new sources are at high $z$.  Our analysis spanning several Spitzer datasets allows us to narrow the estimated contributions of remaining known galaxies to the CIB anisotropies to be probed potentially from surveys by new and upcoming space missions such as Euclid, SPHEREx, and Roman. Of these, the Roman surveys have the best prospects for measuring the source-subtracted CIB and probing the nature of the underlying new populations at $\lambda <2\ \mu$m, followed by Euclid's surveys, while for SPHEREx the source-subtracted CIB signal from them appears significantly overwhelmed by the CIB from remaining known galaxies.
\end{abstract}


\section{Introduction}
The Cosmic Infrared Background, or CIB, in the near-IR (NIR) is the integrated emission due to stars and black hole accretion spanning all redshifts. Uncovering the source density and spatial distribution of galaxies imprinted in the CIB and/or its anisotropies can reveal interesting physics driving galaxy formation and evolution \citep[see review by][and references therein]{Kashlinsky2005review}. Due to the presence of bright but relatively smooth Galactic foregrounds, primarily the Zodiacal light and the interstellar medium (ISM, or Galactic Cirrus), it was suggested to augment direct measurements of the diffuse CIB with its angular anisotropies which are measured via the angular power spectrum (or its Fourier transform, the correlation function) \citep{Kashlinsky1996a,Kashlinsky1996b,Kashlinsky2000}. It was then proposed that anisotropies produced by the clustering of early populations, which are not accessible by direct telescopic studies, could be probed after removing contributions from observed known galaxies \citep{Cooray2004,Kashlinsky2004}, the remaining signal now referred to as the source-subtracted CIB. The subsequent Spitzer-based analysis of the source-subtracted CIB by \cite{Kashlinsky2005} revealed a statistically significant fluctuation component at large (sub-degree) scales which are not accounted for by the remaining known (``ordinary") galaxy populations. 

The uncovered CIB anisotropy signal was confirmed in later Spitzer \citep{Kashlinsky2007data,Kashlinsky2012,Cooray2012,Kashlinsky2025} and AKARI \citep{Matsumoto2011} measurements and was shown to be uncorrelated with visible HST-observed galaxies to $m_{\rm AB}> 28$ \citep{Kashlinsky2007}.
The interpretation of this signal in terms of new populations has been hotly debated, with the current proposals being (1) Population III and early star systems at high-$z$ in the $\Lambda$CDM concordance model  \citep{Kashlinsky2005}, (2) faint galaxies at low-$z$ \citep{Thompson2007a,Thompson2007b}, (3) stellar emission on the outskirts of dark matter (DM) halos from galaxy mergers at $z\sim 1-4$, or Intra-Halo light (IHL) \citep{Cooray2012}, (4) early ($z\geq 10-12$) formation of direct-collapse black holes producing the CIB fluctuation signal \citep{Yue2013,Ricarte2019}, and (5) early star formation increased due to the LIGO-type primordial black holes (PBHs) adding to the inflationary matter power by their inevitable granulation component \citep{Kashlinsky2016,Kashlinsky2025}, the power component first proposed by \cite{Meszaros1974,Meszaros1975}. Furthermore, any interpretation of this CIB fluctuation signal must be able to account for its coherence with the soft X-ray background (CXB) anisotropies \citep{Cappelluti2013,Helgason2014,Mitchell-Wynne2016,Li2018,Cappelluti2017,Kaminsky2025} and very generally implies that the CIB fluctuation must originate in populations containing a much larger proportion of black holes than the presently known populations. The observed CXB-CIB coherence is clearly in conflict with the IHL origin hypothesis. See \cite{Kashlinsky2018} for review. 

In order to more robustly isolate the signal from new populations and accurately discern its properties, understanding the contribution of faint galaxies below the current detection limit of Spitzer is crucial. %\citep{Cooray2004,Kashlinsky2004,Kashlinsky2005review}.
Additionally, the CIB anisotropies originating from faint galaxies may unveil interesting physical connections between them and their host DM halos. This link is also known as the galaxy-halo connection and can be described by the stellar-to-halo-mass relation (SHMR) (e.g., \citealt{Moster2013}; \citealt{Behroozi2019}). Many groups have used the two-point correlation function (or its Fourier inverse, the angular power spectrum) to quantify the SHMR in the stellar mass and redshift ranges of $M_{\star}\sim 10^8-10^{12}$ $M_{\odot}$ and $z\sim 0-15$ respectively (e.g., \citealt{Coupon2012}; \citealt{MartinezManso2015}; \citealt{Hatfield2016}; \citealt{Harikane2016}; \citealt{Shuntov2025shmr}; \citealt{Paquereau2025}; \citealt{Chaikin2025}). Such studies find a peak in the SHMR (denoted as $M_{\rm peak}$), where galaxies in DM halos smaller than $M_{\rm peak}$ undergo suppressed star formation activity due to stellar winds and supernovae feedback while galaxies occupying halo masses greater than $M_{\rm peak}$ have their star formation suppressed by outflows from active galactic nuclei (AGNs) (e.g., \citealt{Lagos2018}; \citealt{ManBelli2018}; \citealt{Contini2025}; \citealt{Chaikin2025}). It has been shown that $M_{\rm peak} \sim 10^{12}$ $M_{\odot}h^{-1}$ for $z\lesssim 4$, while it is expected to increase at higher redshifts (e.g., \citealt{Behroozi2013_lackshmr}; \citealt{Behroozi2013}; \citealt{Ishikawa2020}; \citealt{Paquereau2025}; \citealt{Chaikin2025}).

New measurements of the source-subtracted CIB fluctuations using new Spitzer observations (referred to as ``Looking at infrared background radiation anisotropies with Spitzer", or LIBRAS) have confirmed yet again the large-scale signal that is consistent with a cosmological origin from new populations  \citep{Kashlinsky2025}, which was identified with higher precision and to larger scales. These recent results, in addition to previous measurements of the source-subtracted CIB from \citeauthor{Kashlinsky2007} (\citeyear{Kashlinsky2007}; hereafter KAMM4) and \citeauthor{Kashlinsky2012} (\citeyear{Kashlinsky2012}; hereafter K12), also show a clear small-scale structure that is not consistent with white noise, suggesting a signal due to non-linear galaxy clustering (i.e. galaxies occupying the same DM halos). Using all of these Spitzer-based measurements, the galaxy-halo connection can be studied via Fourier analysis with unprecedented precision.

In this work, we bracket the clustering properties of known galaxy populations contributing to the source-subtracted CIB fluctuations using the new (and old) Spitzer-based measurements at 3.6 and 4.5 $\mu$m, isolating its power components. The remaining known galaxy populations explain the observed small-scale fluctuations to better than a few percent in power meaning that the new populations cannot contribute more than that fraction. This is easy to explain if the new populations originate at high-$z$, but appears in conflict with the IHL model; see the summary of theoretical expectations in Fig. 22 of \cite{Kashlinsky2018}. Using these constraints on the contributions from faint, normal galaxies we address the prospects of CIB-based searches for the new populations with current space missions, Euclid, Roman, and SPHEREx. Of these only Euclid and Roman have good prospects for probing the Spitzer-found CIB from new populations, whereas that signal is significantly subdominant in the SPHEREx surveys because of the much shallower integrations and lower angular resolution.

This paper is organized as follows. We briefly describe the observations, source-clipping procedure, and angular power spectrum formalism in \S 2. In \S 3, we present the theoretical framework used to model the CIB anisotropies. In \S 4, we show the methodology used to constrain the clustering profile of known galaxies and validate our findings with previous works. We then look to measurements with other space telescopes at shorter NIR wavelengths in \S 5 followed by a summary of our results in \S 6. We adopt the cosmological parameters $\Omega_{\rm m}= $ 0.3111, $H_0 = $ 67.66 km s$^{-1}$ Mpc$^{-1}$ from the Planck 2018 results \citep{Aghanim2020}. All magnitudes used in this work are in the AB system \citep{Oke1983}.

\section{Datasets and Fourier Analysis}\label{sec:data}
We employ a suite of Spitzer-based measurements (see Table \ref{tab:Spitzer_fields}) which allow us to probe the source-subtracted CIB anisotropies as a function of the shot noise from remaining galaxies. With the exception of the LIBRAS data, the angular power spectra have been the focus of previous works (see K12 and KAMM4), in which map-making procedures and foreground estimates have already been addressed. All of the datasets are reduced following the same general pipeline \citep{Arendt2010} and are summarized in Table \ref{tab:Spitzer_fields}. In the following subsections we briefly describe each Spitzer dataset and then proceed with the angular power spectrum formalism.

\begin{deluxetable*}{llcccc}
\tablecaption{Summary of Spitzer Datasets \label{tab:Spitzer_fields}}
\tablehead{
\colhead{Dataset} & 
\colhead{Region} & 
\colhead{$(\ell, b)$ [deg]} & 
\colhead{Size [arcmin]} & 
\colhead{$\langle t_{\mathrm{obs}} \rangle$ [hr]} & 
\colhead{$f_{\rm sky}$}
}
\startdata
LIBRAS & NEP        & (95.8, 29.9)       & $81.9 \times 81.9$   & $\sim 1.8$ & 0.61 \\
       & CDFS       & (224.1, $-54.6$)   & $124 \times 57$      & $\sim 1.8$ & 0.63 \\
SEDS   & UDS        & (169.98, $-59.88$) & $21 \times 21$       & 13.6 & 0.730 \\
       & EGS        & (95.95, 59.81)     & $8 \times 62$        & 12.5 & 0.725 \\
GOODS  & HDF-N E1   & (125.9, 54.8)      & $10.2 \times 10.2$   & 20.9  & $\geq 0.65$ \\
       & HDF-N E2   & (125.9, 54.8)      & $10.2 \times 10.2$   & 20.7   & $\geq 0.65$ \\
       & CDF-S E1   & (223.6, $-54.4$)   & $8.8 \times 8.4$     & 23.7   & $\geq 0.65$ \\
      & CDF-S E2   & (223.6, $-54.4$)   & $9.0 \times 8.4$     & 22.4   & $\geq 0.65$ \\
\enddata
\tablecomments{Parameters of each Spitzer dataset are shown and are sorted according to depth from top-to-bottom, with LIBRAS being the shallowest and GOODS being the deepest.}
\end{deluxetable*}
\subsection{LIBRAS Data}
\cite{Kashlinsky2025} probed the CIB anisotropies to angular scales previously unobtained ($\sim 1^{\circ}$). As described in \cite{Kashlinsky2025}, subsets from public data of the Spitzer IRAC 3.6 and 4.5 $\mu$m observations from Program IDs 13153 and
13058 \citep{Capak2016} are utilized in the north ecliptic pole (NEP) and Chandra Deep Field South (CDFS) fields. The NEP field corresponds to the Euclid Deep Field North, while CDFS matches the Euclid Deep Field Fornax.

\begin{figure*}[ht]
    \centering
    \includegraphics[width=\textwidth]{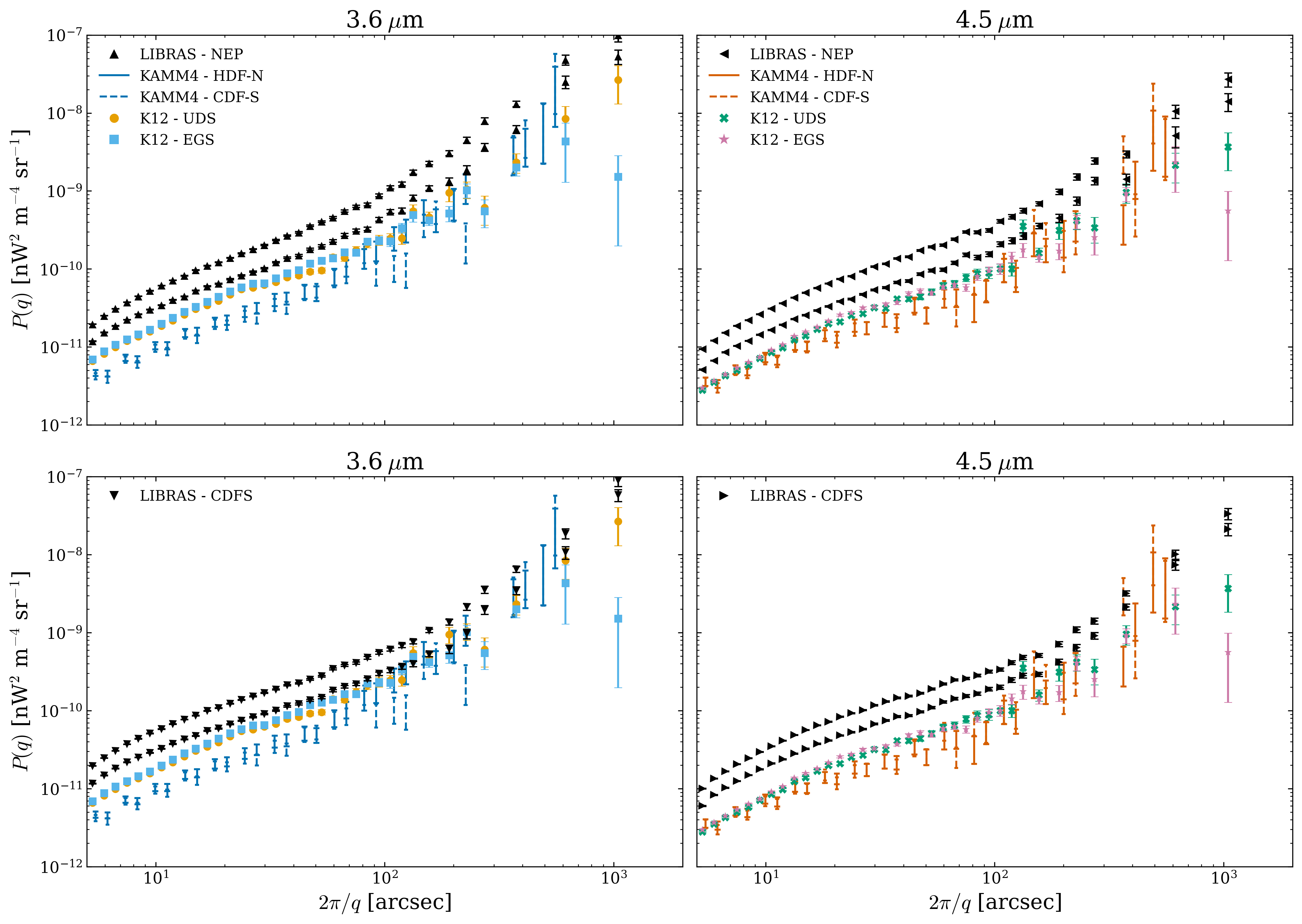}
    \caption{Angular power spectra computed from three separate Spitzer datasets are shown. The gray points mark the new LIBRAS data \citep{Kashlinsky2025}. The red and blue error bars mark the K12 data from the UDS and EGS fields respectively (K12). The green and orange data points represent the HDF-N and CDF-S data over two observational epochs (E1 and E2) respectively (KAMM4). KAMM4 points for each epoch are shifted for clarity. Errors are reported at the $1\sigma$ confidence level.}
    \label{fig:cib_power_spectra_data}
\end{figure*}
\subsection{SEDS Data}
The second dataset we use is introduced and analyzed by K12, and comes from the Spitzer Extended Deep Survey (SEDS) program \citep{Fazio2011}. The two SEDS fields studied were the Ultra-Deep Survey field (UDS) and the Extended Groth Strip (EGS) and observed at three observational epochs separated by $\sim6$ months. 

The angular power spectra at each epoch are cross-correlated to test that the observed signals are not due to the detector and/or zodiacal light (i.e. that the signal is extragalactic in origin), and indeed show that such contributions are negligible (see Fig. 4 of K12). As a result, in this work we analyze the angular power spectra computed over all observational epochs in each SEDS field. For further detail on the data reduction we refer the reader to K12.
\subsection{GOODS Data}
The deepest observations of the source-subtracted CIB made thus far were in the Great Observatories Origins Deep Survey (GOODS; \citealt{Dickinson2003}). We describe briefly the datasets here and refer the reader to KAMM4 and \cite{Arendt2010} for further details and methodology.  

The chosen GOODS fields were Hubble Deep Field North (HDF-N) and Chandra Deep Field South (CDF-S; a portion of the field not covered by the LIBRAS data above). Furthermore, observations were made at two epochs as in \cite{Kashlinsky2012}. The exposures in both maps are very deep at $\sim 20$ hours per pixel corresponding to shot noise levels of $\sim 1-2$ $\times$ $ 10^{-11} $ nW$^2$ m$^{-4}$ sr$^{-1}$ for both the 3.6 and 4.5 $\mu$m maps. Because uniform shot noise levels are reached across both maps and observational epochs, we stack all of the angular power spectra.

\subsection{Fourier Analysis}
Each CIB map is initially self-calibrated using the procedure from \cite{Fixsen2000}. Resolved sources (e.g. stars, galaxies, etc.) are masked using an iterative procedure where the flux at a given pixel $F_i$ is removed under the condition that $F_i \geq \langle F \rangle + N_{\rm cut}\sigma_F$ with $N_{\rm cut} = 3$. Additionally, 3 $\times$ 3 pixels are removed to account for the beam. For each field, the $3.6$ and $4.5$ $\mu$m masks are joined together and applied to both maps. For a robust Fourier analysis the fraction of unmasked pixels $f_{\rm sky}$ should be $\geq 60\%$. After masking, an iterative procedure was applied to remove the extended wings of both resolved and unresolved sources which is a variant of the CLEAN algorithm  \citep{Hogbom1974} (see \cite{Arendt2010} for further detail). The LIBRAS and SEDS maps were constructed with a 1$''$.2 pixel scale, while the GOODS data with a finer 0$''$.6 pixel scale. While we only evaluate the angular power spectrum from K12 and KAMM4 at a single iteration of the source model (previously selected in each respective work), we evaluate two iterations of the LIBRAS source-subtracted maps in order to investigate a wider magnitude depth range. 

Each of these masked, source-subtracted maps were converted into fluctuation fields $\delta F(\vec{x})$. The Fourier transform
\begin{equation}
\Delta (\vec{q}) = \int \delta F(\vec{x}){\rm exp}({-i\vec{x}\cdot \vec{q}})d^2x
\end{equation}
is then computed, after subtracting the mean background $\langle F\rangle$, where $q$ is the spatial frequency ($\theta =2\pi /q$). For each $q$-ring given by [$q$, $q+\delta q$], the azimuthal average over the ring containing $N_q$ independent Fourier elements was computed to determine the angular power spectrum
\begin{equation}
P(q) = \frac{\langle {| \Delta (\vec{q}) |}^2 \rangle_{q\rm-ring}}{f_{\rm sky}}
\label{eq:ap}
\end{equation}
where the $1/f_{\rm sky}$ factor was included to correct for masking pixels in the image. The uncertainty is calculated as $\sigma(q)=P(q)/\sqrt{N_q}$, where Poisson statistics are assumed to account for the cosmic variance \citep{Abbott1984}. Alternate methods of characterizing cosmic variance in the context of number counts and integrated galactic light (IGL) are discussed by \cite{Tompkins2025} and \cite{Carter2025}. For the background power spectrum evaluation, noise maps ($A-B$ for independent $A$, $B$ subsets) were generated \citep{Kashlinsky2005,Arendt2010,Kashlinsky2007,Kashlinsky2012,Kashlinsky2025}. The noise power measured in the $A-B$ maps is subtracted from the overall $P(q)$ and the cosmic variance uncertainty propagated, leading to the finalized angular power spectra shown in Fig. \ref{fig:cib_power_spectra_data}.

\section{Modeling the CIB Anisotropies from known galaxies}
In order to reconstruct the contributions of known galaxy populations to the source-subtracted CIB fluctuations, we must be able to account for their flux distribution and how they are clustered in large-scale structures \citep{Kashlinsky2005}. While we retain the reconstructed flux distributions and galaxy counts from HRK12 (see below), we provide a modified halo model using newer galaxy clustering measurements that extend to higher redshifts.

\subsection{Source Densities and Flux Production Rates}
To reconstruct the the source density of remaining galaxies contributing to the source-subtracted CIB, we use the empirical galaxy population model of HRK12. This model populates the lightcone using galaxy luminosity functions (LFs) of the Schechter form \citep{Schechter1976} in the \textit{rest-frame} UV/Optical/NIR over redshifts $0<z<8$. The evolution of the Schechter parameters $\phi^*$, $M^*$, and $\alpha$ (corresponding to the LF normalization, characteristic absolute magnitude, and faint-end slope respectively) are fitted as a function of redshift and interpolated over the rest-frame LFs to determine the number counts in a range of \textit{observed} NIR wavelengths. See HRK12 for further information.

\begin{figure*}[ht]
    \centering
    \includegraphics[width=\textwidth]{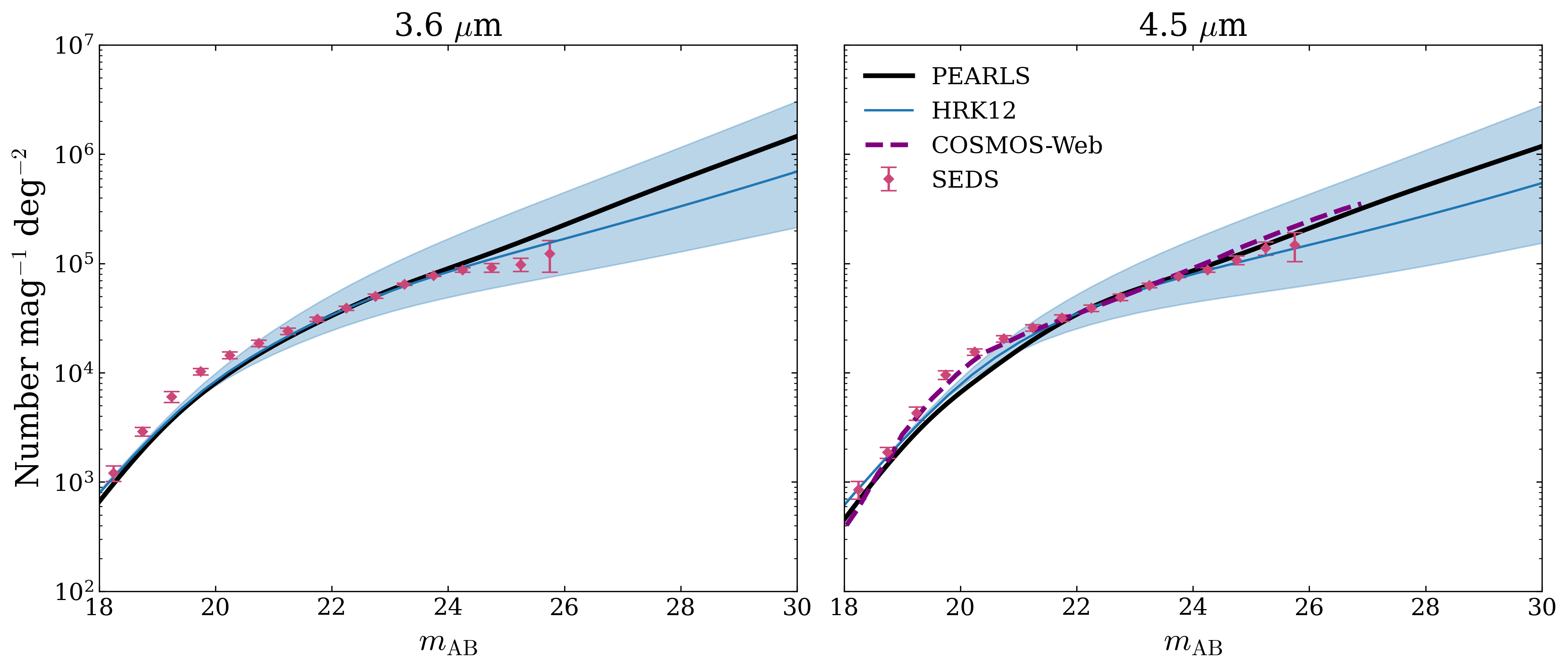}
    \caption{Galaxy counts are shown at $m_{\rm AB} \gtrsim 18$. The shaded blue region spans the LFE and HFE models with the solid blue line being the DFE model, all of which are obtained using the best-fit Schechter parameter evolution fits from HRK12. The red diamonds indicate Spitzer SEDS galaxy counts from \cite{Ashby2013}. The dashed purple and solid black lines are galaxy counts from the JWST COSMOS-Web and PEARLS surveys respectively \citep{Windhorst2023,Shuntov2025cat}. Errors are reported at the $1\sigma$ confidence level.}
    \label{fig:number_counts_comparison}
\end{figure*}

\begin{figure*}[ht]
    \centering
    \includegraphics[width=\textwidth]{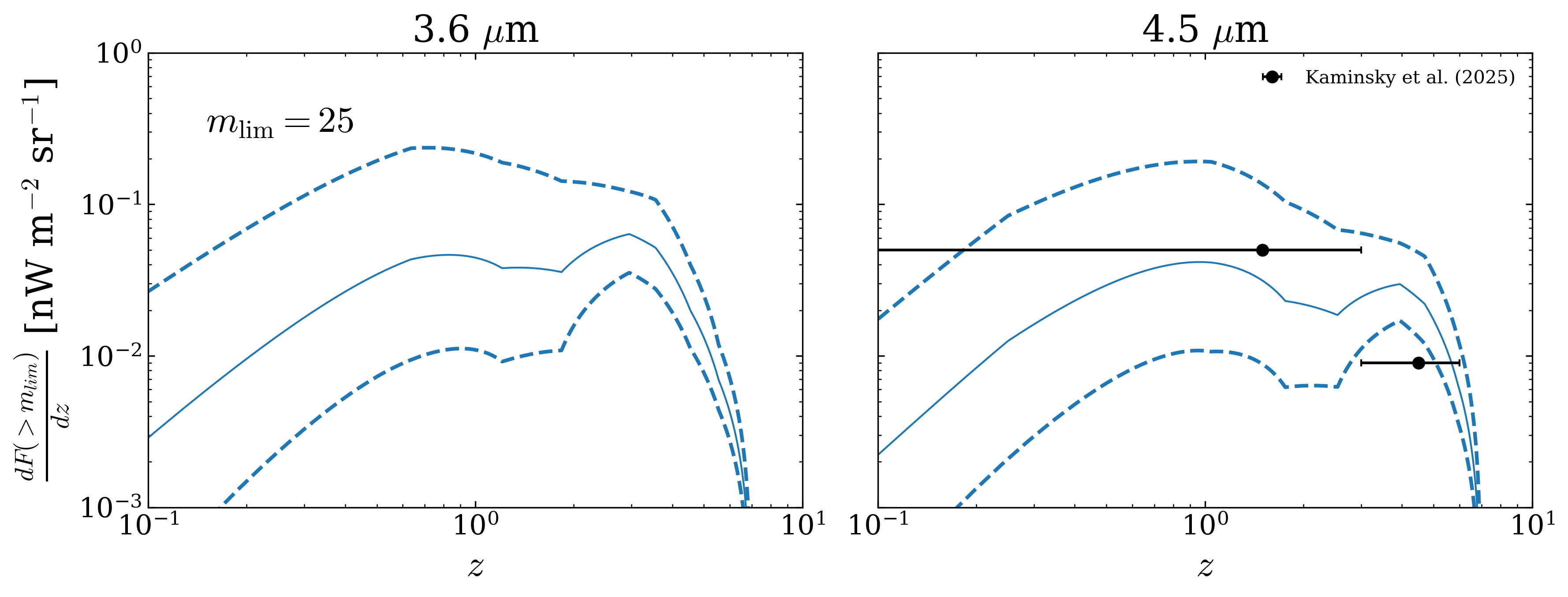}
    \caption{Flux production rates are shown for $m_{\rm lim}=25$ and $z_{\rm max}=7$ (as chosen in HRK12). The lower and upper dashed lines represent the LFE and HFE limits of the galaxy reconstruction respectively, and the thin solid line represents the DFE, or default model. The black data points show the flux production rates computed from galaxies resolved by JWST which are $25 \leq m_{\rm AB}\lesssim 29$ \citep{Kaminsky2025}. There is broad agreement between the empirical model and the results from \cite{Kaminsky2025}. }
    \label{fig:dFdz_mlim_25_comp}
\end{figure*}

Due to less data on LFs at longer rest-frame wavelengths (and hence lower statistics), there is an allowed spread in the evolution of $\alpha$, with its lower and upper limits referred to as the low faint-end (LFE) and the high faint-end (HFE) respectively \citep[but see Fig. 2 in][]{Helgason2012b}. The LFE and HFE models bracket the true source densities in each NIR band. The galaxy counts in a given NIR wavelength for either faint-end limit can be written as 
\begin{equation}
N(m_{\rm AB})=\int_{z_{\rm min}}^{z_{\rm max}}dz \frac{dV}{dzd\Omega} \Phi_{\mathrm{obs}}(m_{\rm AB}|z)
\end{equation}
where $\Phi_{\rm obs}$ is the interpolated LFE/HFE LF and $dV/dzd\Omega$ is the co-moving volume element. As in HRK12 we take $z_{\rm min}=0$ and $z_{\rm max}=7$. Furthermore, we can also compute the flux production rate from sources fainter than a given AB magnitude $m_{\rm lim}$
\begin{equation}
\frac{dF(>m_{\rm lim})}{dz}=\int_{m_{\textrm{lim}}}^\infty dm_{\rm AB}f(m_{\rm AB})\frac{dN(m_{\rm AB}|z)}{dz}
\label{eq:flux_prod_rate}
\end{equation}
where $f(m_{\rm AB})=\nu f_\nu$ and $f_\nu = 10^{-0.4(m_{\rm AB}+33.6)}$ nW m$^{-2}$ Hz$^{-1}$. The average of the LFE and HFE limits is taken as the ``default" faint-end (DFE) model. As seen in Figs. \ref{fig:number_counts_comparison} and \ref{fig:dFdz_mlim_25_comp}, the empirical model from HRK12 is highly consistent with galaxy counts from Spitzer \citep{Ashby2013} and JWST \citep{Windhorst2023,Shuntov2025cat}.
%This methodology has been validated with further galaxy surveys, observed with both Spitzer and JWST.

\subsection{CIB Fluctuations}\label{subsec:halo_models}   
With the HRK12 empirical reconstruction, we can move forward with modeling the CIB angular power spectrum. The first contribution to the power spectrum is the shot noise component which is generated by galaxies entering the beam, and as a result behaves as a white noise spectrum \citep{Kashlinsky2005review,Kashlinsky2012,Cooray2012,Kaminsky2025}. The shot noise can be written as
\begin{equation}
P_{\rm SN}=\int_{z_{\rm min}}^{z_{\rm max}}dz \int_{m_{\rm lim}}^{\infty}dm_{\rm AB}f^2(m_{\rm AB})\frac{dN(m_{\rm AB}|z)}{dz}
\label{eq:shot_noise}
\end{equation}
which is scale-independent outside the beam. 

Once the shot noise component is treated we can focus on the fluctuation signal originating from the spatial distribution of galaxies. In the $\Lambda$CDM cosmology, density perturbations $\delta \rho$ evolve into objects such as the galaxies and galaxy clusters that we observe today. On larger scales $\delta \rho<1$ and are well described by the linear approximation. However, on smaller spatial scales perturbations $\delta \rho \gg 1$ and grow out of the linear regime. In order to account for galaxy clustering in the linear \textit{and} non-linear regimes, we use the halo model described by \cite{CooraySheth2002}. In this formalism, the 3D power spectrum is given by
\begin{equation}
P_{\rm 3D}(k,z)=P^{\rm 1h}(k,z)+P^{\rm 2h}(k,z)
\end{equation}
where the $P^{1\rm h}$ (1-halo) term accounts for the clustering of galaxies in the same DM halo, while the $P^{2\rm h}$ (2-halo) term accounts for galaxies that occupy separate halos. According to \cite{SkibbaSheth2009}, each of these components can be written as
\begin{equation}
P^{\rm{1h}}(k,z) =
\int dM_{\rm h} \frac{dn}{dM_{\rm h}} \langle N_{\rm{cen}} \rangle\left( 
\frac{2\langle N_{\rm{sat}} \rangle u_{\rm NFW}(k|M_{\rm h},z)+\langle N_{\rm{sat}} \rangle ^2 u_{\rm NFW}^2(k|M_{\rm h},z)}{{ \bar{n}_{\rm{gal}}}^2}
\right)
\label{eq:1halo}
\end{equation}

and
\begin{equation}
P^{\rm{2h}}(k,z)=P^{\rm{lin}}(k,z)
\times \left( \int dM_{\rm h}\frac{dn}{dM_{\rm h}}\langle N_{\rm{cent}}\rangle\frac{1+\langle N_{\rm{sat}}\rangle u_{\rm NFW}(k|M_{\rm h},z)}{ \bar{n}_{\rm{gal}}}b(M_{\rm h},z) \right)^2
\label{eq:2halo}
\end{equation}
respectively, where $dn/dM_{\rm h}$ is the halo mass function (HMF, which we take from \citealt{Behroozi2013}), $u_{\rm NFW}(k|M_{\rm h},z)$ is the Fourier-transform of the Navarro-Frenk-White (NFW) density profile \citep{NFW1996}, and $b(M_{\rm h},z)$ 
is the linear bias adopted from \cite{Sheth2001}. When implementing the NFW profile we must also assume a concentration-mass relation $c(M_{\rm h},z)$. We compute this relation using Eq. 4 of \cite{Duffy2008} where 
\begin{equation}
c(M_{\rm h},z)=\frac{A}{(1+z)^C}\left(\frac{M_{\rm h}}{M_{\rm pivot}}\right)^B
\end{equation}
and adopt the parameters $A=9$, $B=-0.13$, and $C=1$ \citep{CooraySheth2002,Amblard2011}, and take the pivot mass $M_{\rm pivot}$ to be the characteristic mass scale at which the peak height $\nu=\delta_c/\sigma(M_{\rm h})$ equals unity at a given $z$. We discuss the sensitivity of our models to our chosen HMF and $c(M_{\rm h},z)$ parameterizations in \S \ref{subsec:syst_unc}. 

The last three terms from Eqs. \ref{eq:1halo} and \ref{eq:2halo} that we need to define are $\langle N_{\mathrm{cent}}\rangle$, $\langle N_{\mathrm{sat}}\rangle$, and $\bar{n}_{\rm{gal}}$, which represent the mean occupation number of central and satellite galaxies and the mean total galaxy density respectively. These quantities are computed using the Halo Occupation Distribution (HOD) statistical framework, which quantifies how the mean number of central and satellite galaxies occupying DM halos change as a function of halo mass. In this study we use the five-parameter model introduced by \cite{Zheng2005}, where the mean occupation numbers of central and satellite galaxies can be written as
\begin{equation}
\langle N_{\mathrm{cent}}\rangle=\frac{1}{2}\left[1+\mathrm{erf}\left(\frac{\mathrm{log}(M_{\rm h})-\mathrm{log}(M_{\mathrm{min}})}{\sigma_{\mathrm{log}M_{\rm h}}}\right)\right]
\label{eq:N_cent}
\end{equation}
and 
\begin{equation}
\langle N_{\mathrm{sat}}\rangle=\left(\frac{M_{\rm h}-M_0}{M_\mathrm{sat}}\right)^{\alpha_{\rm sat}}
\label{eq:N_sat}
\end{equation}
respectively. In Eq. \ref{eq:N_cent}, $M_{\rm min}$ corresponds to the minimum halo mass at which $\langle N_{\mathrm{cent}}(M_{\rm min})\rangle=0.5$ and $\sigma_{{\rm log}M_{\rm h}}$ corresponds to the step width for a halo having one central galaxy. Additionally, as seen in Eq. \ref{eq:N_sat}, $M_0$ is the mass threshold for which a halo can host a satellite galaxy, $M_{\rm sat}$ is the characteristic mass at which a halo hosts a satellite galaxy, and $\alpha_{\rm sat}$ controls how abundant satellites are for a given halo mass. It has been shown that $M_{\rm sat}$ and $M_0$ can be related with the following relation
\begin{equation}
{\rm log}(M_0/{M_{\odot}h^{-1}})=0.76\times{\rm log} \left(M_{\rm sat}/{M_{\odot}h^{-1}}\right)+2.3
\label{eq:M0_Msat}
 \end{equation}
which has been verified at $z\lesssim3$, although previous studies extrapolate this relation to higher redshifts \citep{Conroy2006,Harikane2016,Paquereau2025}. We can further further parameterize $M_{\rm sat}$ by introducing the quantity $\Delta_{\rm sat}$ where $M_{\rm sat}=10^{\Delta_{\rm sat}}\times M_{\min}$. With these quantities another necessary component is the mean occupation number of \textit{all} galaxies in an individual DM halo, written as
\begin{equation}
\langle N_{\mathrm{gal}}\rangle=\langle N_{\mathrm{cent}}\rangle\times(1+\langle N_{\mathrm{sat}}\rangle).
\label{eq:N_gal}
\end{equation}
The above prescriptions deviate from the original HRK12 formalism, which adopts a four-parameter halo model (see their Eq. 19). Moving to a five-parameter prescription like in this work improves the flexibility of our modeling. Finally, using Eqs. \ref{eq:N_cent} -- \ref{eq:N_gal} we can compute the mean galaxy density as
\begin{equation}
\bar{n}_{\rm gal}=\int dM_{\rm h}\frac{dn}{dM_{\rm h}}(M_{\rm h},z)\langle N_{\mathrm{gal}}\rangle
\label{eq:mean_gal_dens}
\end{equation}
where we set the upper integration bound to $M_{\rm max}=10^{14}$ $M_{\odot}h^{-1}$ which is chosen using the results of \cite{Helgason2017}. This chosen $M_{\max}$ value is also used in Eqs. \ref{eq:1halo} and \ref{eq:2halo}. For the remainder of this study we will express all halo mass terms in the general form $\log(M_{\rm h}/M_{\odot}h^{-1})$.

Putting all of these ingredients together, the 3D galaxy power spectrum is projected into 2D using the Limber approximation \citep{Limber1953}
\begin{equation}
P_{\rm Cl}(q)=\int_{z_{\rm min}}^{z_{\rm max}}dz \frac{H(z)}{cd_c^2(z)}\left(\frac{dF}{dz}\right)^2P_{\rm 3D}(k=qd_c^{-1},z)
\label{eq:norm_gal_clustering}
\end{equation}
where $c$, $d_c(z)$, and $H(z)$ is the speed of light, co-moving distance, and Hubble term respectively. In total, the angular power spectrum of the known galaxy populations can be written as
\begin{equation}
P(q)=P_{\rm SN}+P_{\rm Cl}(q).
\end{equation}
%\section{Constraining the Small-Scale CIB Anisotropies}\label{sec:results}
\section{Understanding the Small-Scale CIB Anisotropies}\label{sec:results}
\subsection{Bracketing Methodology}
\label{subsec:methods}
In this study we seek to constrain the behavior of CIB angular power spectra using the LFE/HFE galaxy counts and HOD statistics.  For each power spectrum computed using Spitzer data (see Fig. \ref{fig:cib_power_spectra_data}), the shot noise $P_{\rm SN}$ is bracketed by $m_{\lim}^{\rm LFE}$ and $m_{\lim}^{\rm HFE}$ (i.e. both magnitude limits will ideally reproduce the same observed shot noise level). For each faint-end model we extend this bracketing to the HOD parameters as well. In this machinery we take the HOD parameters to be redshift-independent. This prescription is supported by galaxy clustering measurements that find mild-to-little evolution as a function of redshift at $1\lesssim z\lesssim 3-4$ (e.g., \citealt{Hatfield2016}; \citealt{Contreras2017}; \citealt{Contreras2023}; \citealt{Paquereau2025}). At higher redshifts the evolution is more noticeable, but most of the NIR emission produced by faint galaxies occurs in the interval $1\lesssim z\lesssim 3$ so any significant evolution at higher redshifts should not meaningfully impact our results. We discuss the sensitivity of this assumption later in this section. 

Furthermore, we opt to fix $\sigma_{{\rm log}M_{\rm h}}$, $\alpha_{\rm sat}$, and $\Delta_{\rm sat}$ for both the LFE and HFE models to avoid degeneracies and prioritize constraining the $\log(M_{\min}^{\rm LFE,HFE}/M_{\odot}h^{-1})$ parameters (e.g., \citealt{Zehavi2011}; \citealt{Harikane2016}; \citealt{Paquereau2025}). Prior galaxy clustering measurements have found degeneracies between the HOD parameters which can be broken using the conditional luminosity function (CLF), which links galaxy luminosity to DM halo mass. It is important to note however that not using the CLF has been found to be a valid approximation when comparing the HRK12 reconstruction to the Millennium Simulation semi-analytical model (SAM) \citep{Guo2011,Helgason2014,Helgason2017}. 
Using a multitude of studies that constrain the HOD statistics as both a function of redshift and stellar mass, we adopt the values $\sigma_{\log M_{\rm h}}=0.2$ dex \citep{Zheng2005,Wake2011,Harikane2016,Paquereau2025}, $\alpha_{\rm sat}=0.75$ \citep{Durkalec2018,Paquereau2025}, and $\Delta_{\rm sat}=0.8$ dex \citep{McCracken2015,MartinezManso2015,Harikane2016,Hatfield2016,Durkalec2018,Paquereau2025}. We discuss the sensitivity of our constraints to adopting these values later in this section.

In total, we bracket $m_{\rm lim}$ and $\log(M_{\rm min}/M_{\odot}h^{-1})$ for each power spectrum (see Fig. \ref{fig:cib_power_spectra_data}) using the following functions
\begin{equation}
P^{\rm LFE}(q)=b_{\rm Spitzer}^2(q)\times[ P_{\rm SN}^{\rm LFE}+P_{\rm Cl}^{\rm LFE}(q)]
\label{eq:lfe_p}
\end{equation}
and
\begin{equation}
P^{\rm HFE}(q)=b_{\rm Spitzer}^2(q)\times[P_{\rm SN}^{\rm HFE}+P_{\rm Cl}^{\rm HFE}(q)]
\label{eq:hfe_p}
\end{equation}
where $b_{\rm Spitzer}(q)$ is the beam transfer function which convolves the true angular power spectrum with the shape of the Spitzer beam. While it is assumed to be Gaussian, there are deviations expected which introduce systematic errors at angular scales below $\sim 10''$. 
\subsection{Results}\label{subsec:results}

\begin{figure*}[ht]
  \centering
  \includegraphics[width=\textwidth]{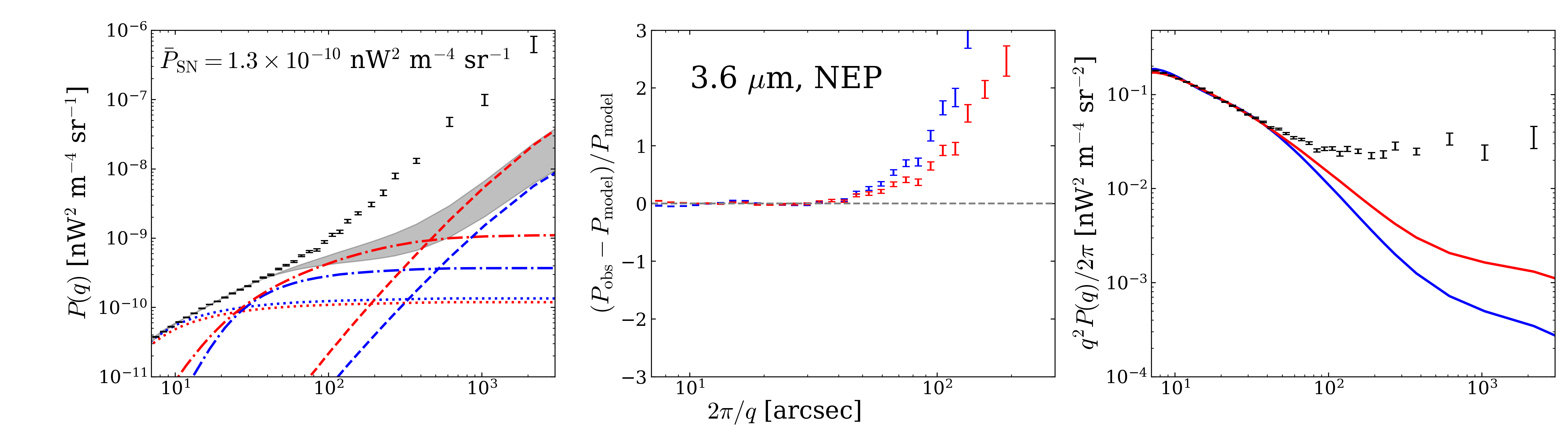}\\[1ex]
  \includegraphics[width=\textwidth]{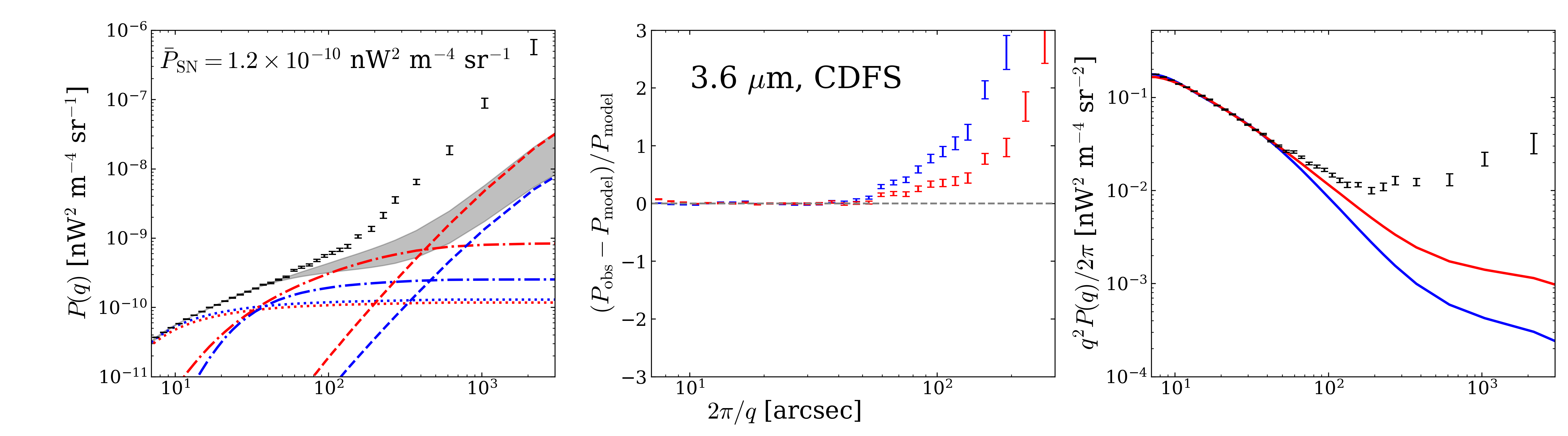}\\[1ex]
  \includegraphics[width=\textwidth]{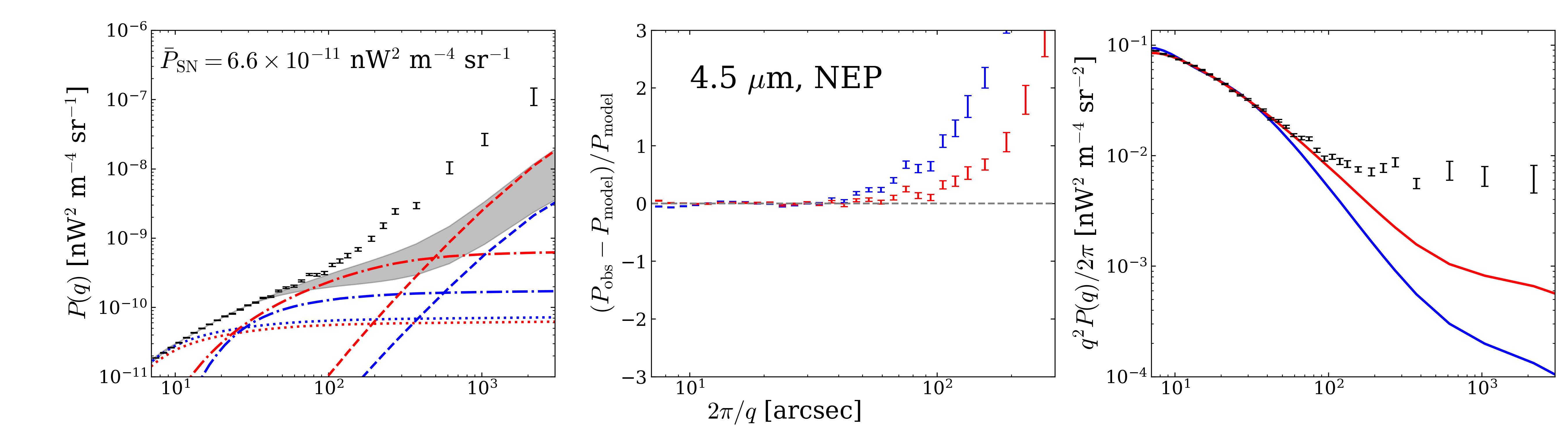}\\[1ex]
  \includegraphics[width=\textwidth]{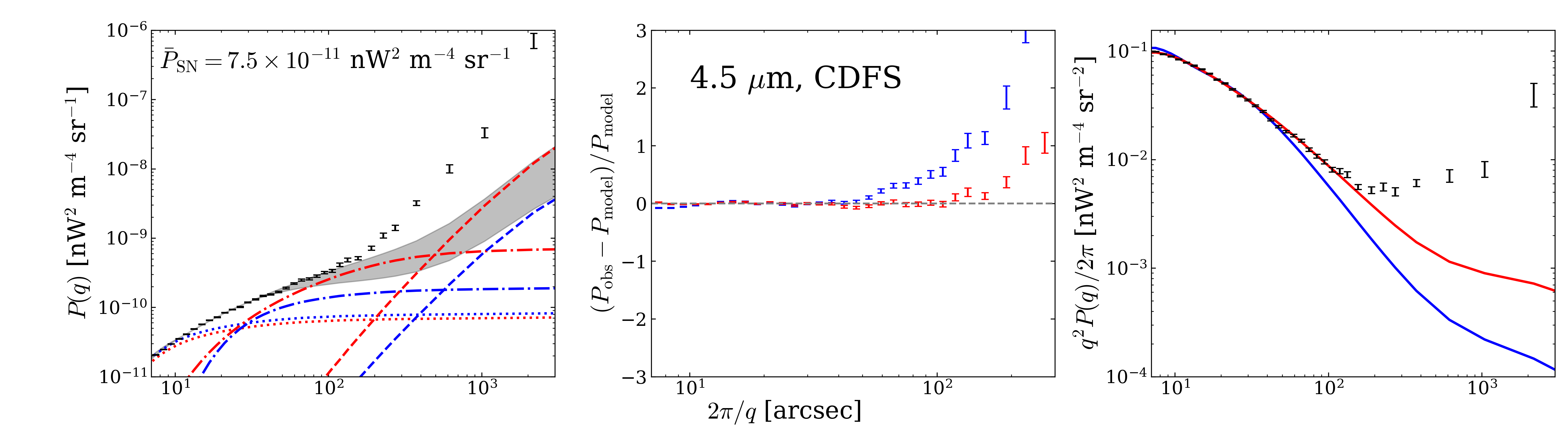}
  \caption{CIB reconstructions using the HRK12 empirical model for each LIBRAS angular power spectrum (represented by the black error bars). The blue lines are of the LFE limit and the red lines the HFE limit. For each FE model, the dotted, dash-dotted, and dashed lines show contributions from the shot noise, 1-halo, and 2-halo power respectively. The shaded gray area represents the allowed range between the LFE and HFE models. The left column shows the power spectrum, the middle column shows the relative error in the model bracketing, and the right column shows the fluctuation spectrum. The average shot noise level $\bar{P}_{\rm SN}$ = mean($P_{\rm SN}^{\mathrm{LFE,HFE}}$) is shown in the left column for each power spectrum. Errors are reported at the 1$\sigma$ level.}%
  \label{fig:LIBRAS_vertical_triple_panels_li}
\end{figure*}

\begin{figure*}[ht]
  \centering
  \includegraphics[width=\textwidth]{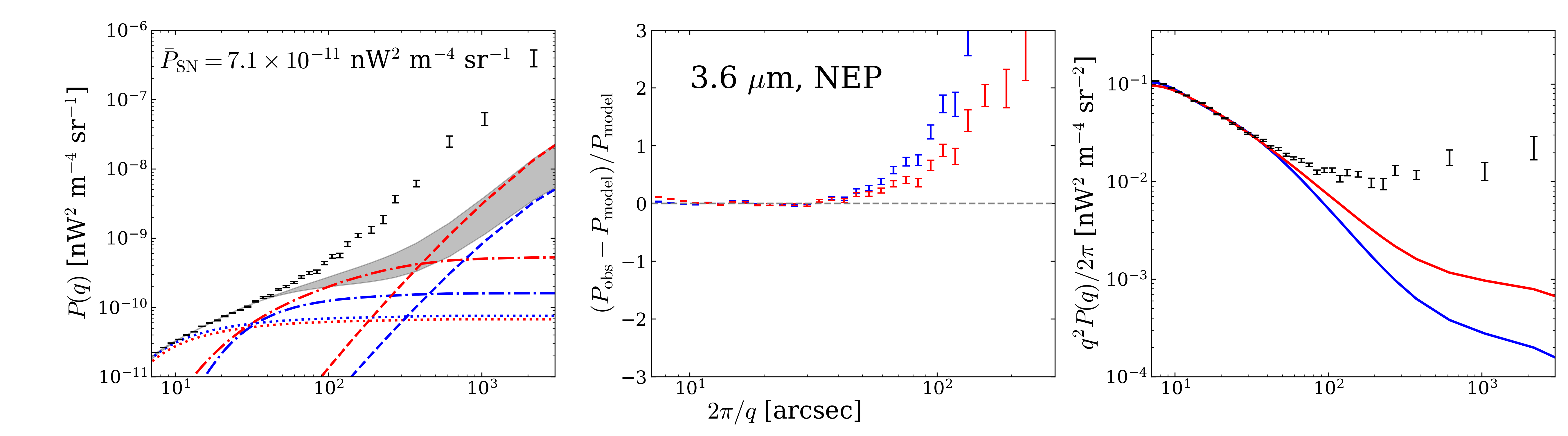}\\[1ex]
  \includegraphics[width=\textwidth]{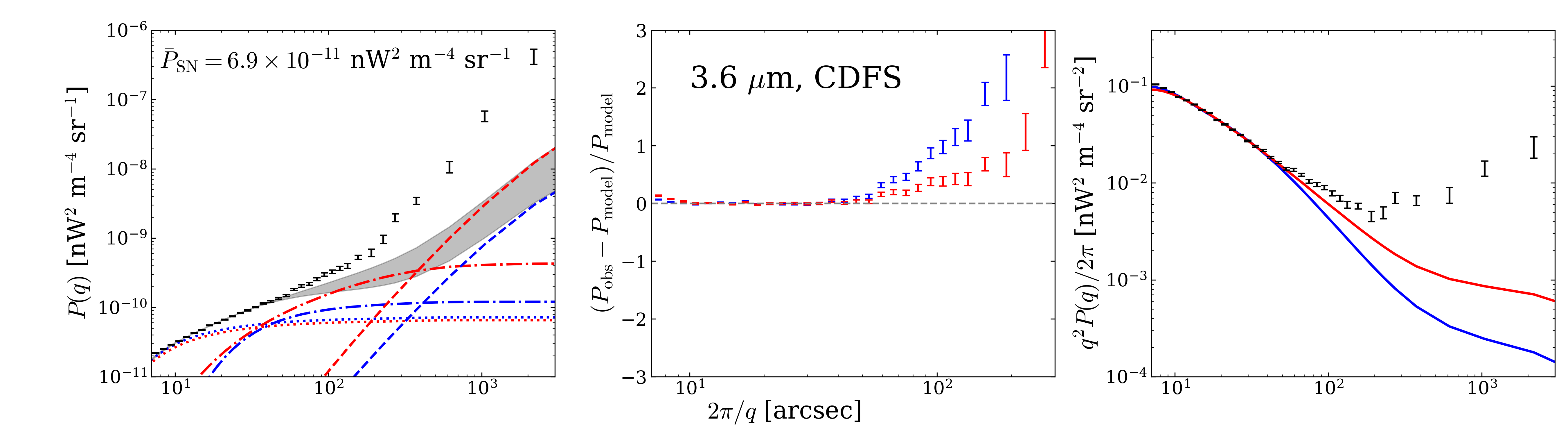}\\[1ex]
  \includegraphics[width=\textwidth]{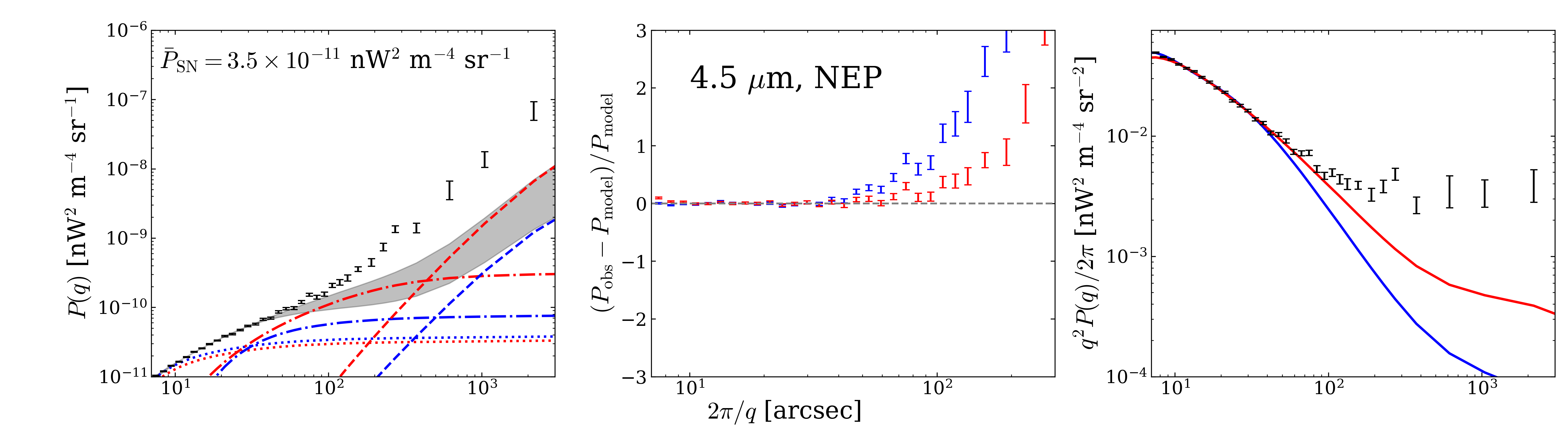}\\[1ex]
  \includegraphics[width=\textwidth]{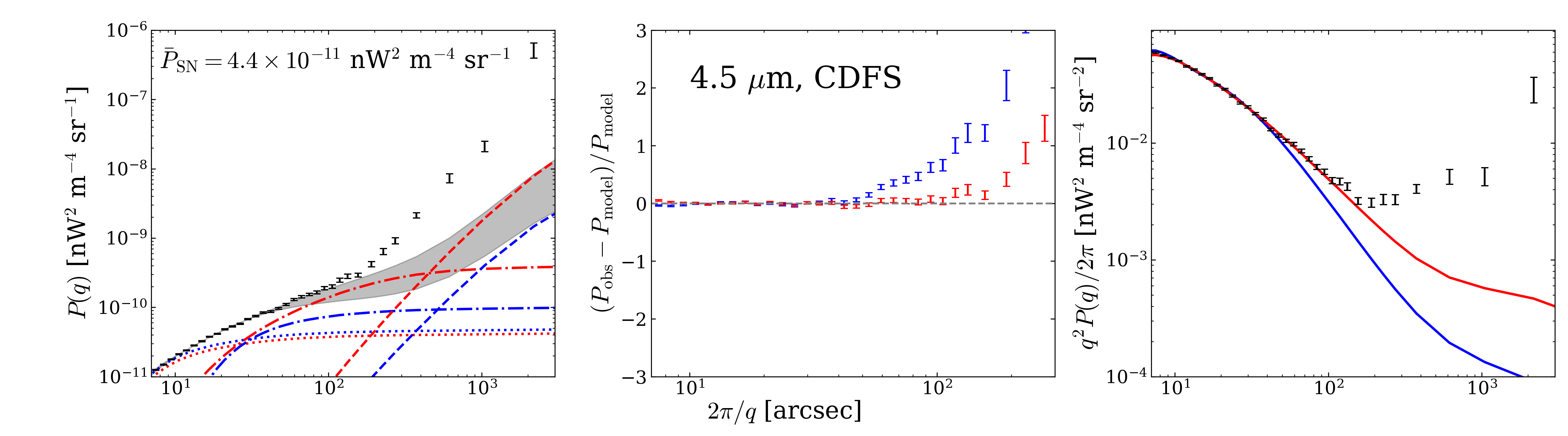}
  \caption{CIB reconstructions using the HRK12 empirical model for each LIBRAS angular power spectrum (represented by the black error bars). The reconstructed power spectra are shown using the same formatting as those in Fig. \ref{fig:LIBRAS_vertical_triple_panels_li}, with the only difference being that the shot noise levels are lower by a factor of $\sim2-3$.}%
  \label{fig:LIBRAS_vertical_triple_panels_hi}
\end{figure*}

\begin{figure*}[ht]
    \centering
    \includegraphics[width=\textwidth]{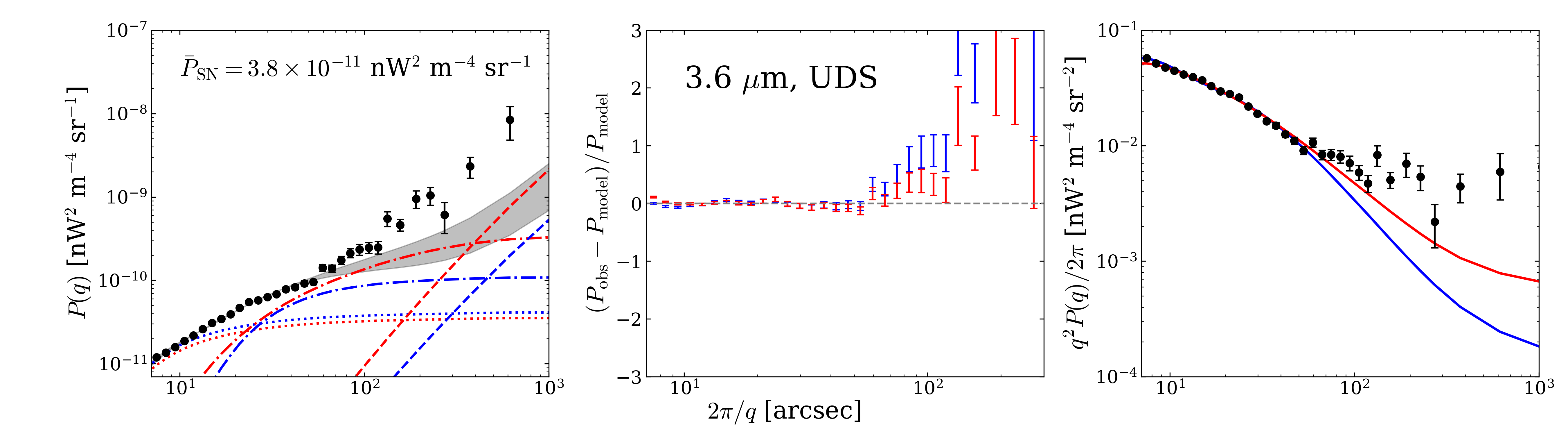}
    \includegraphics[width=\textwidth]{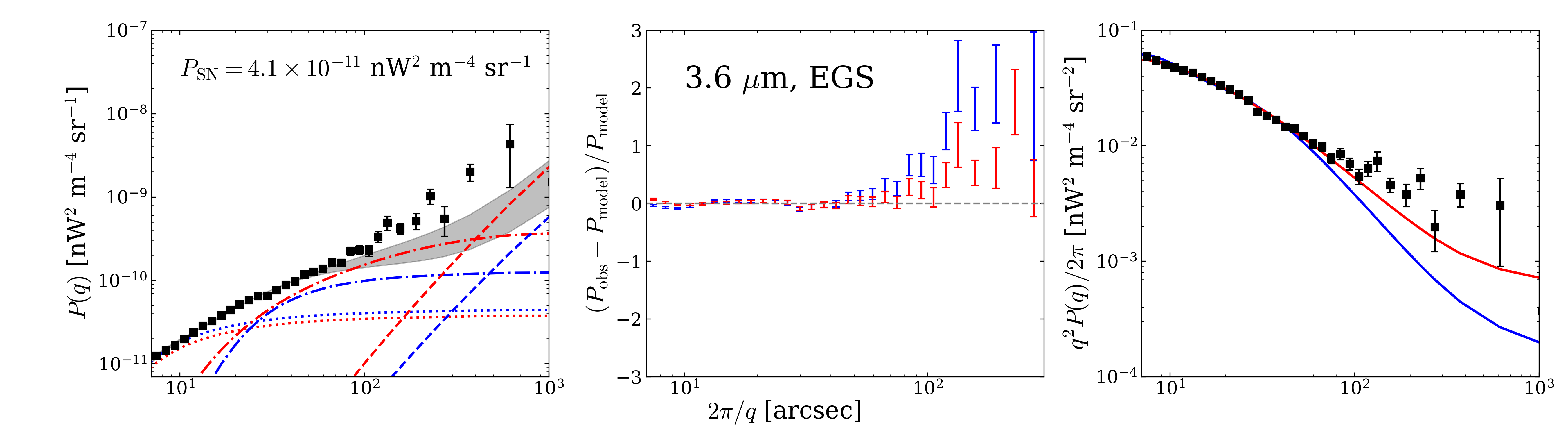}
    \includegraphics[width=\textwidth]{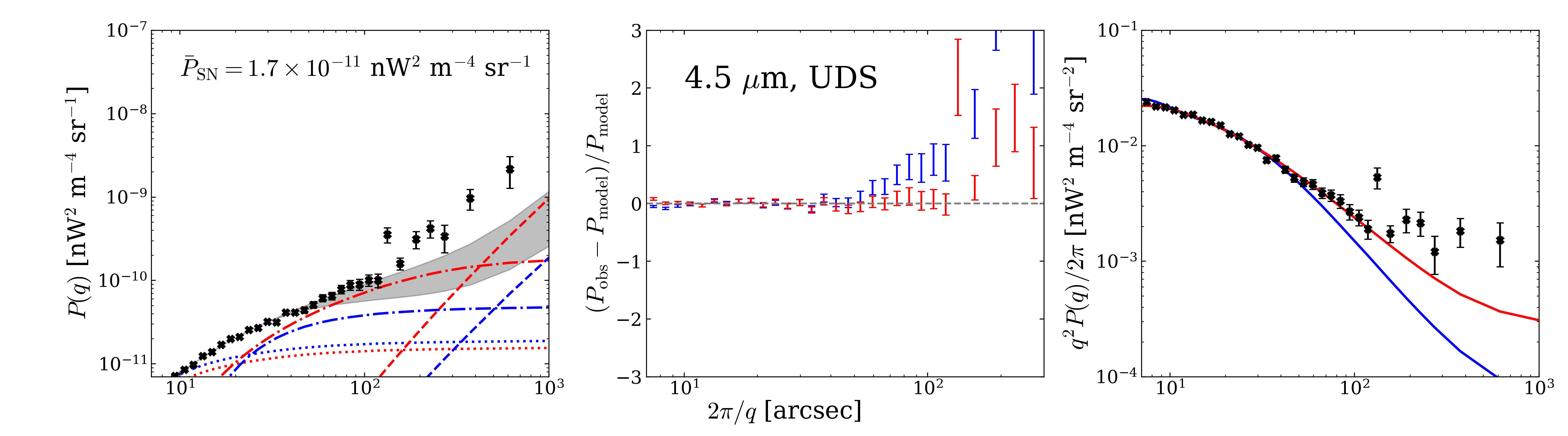}
    \includegraphics[width=\textwidth]{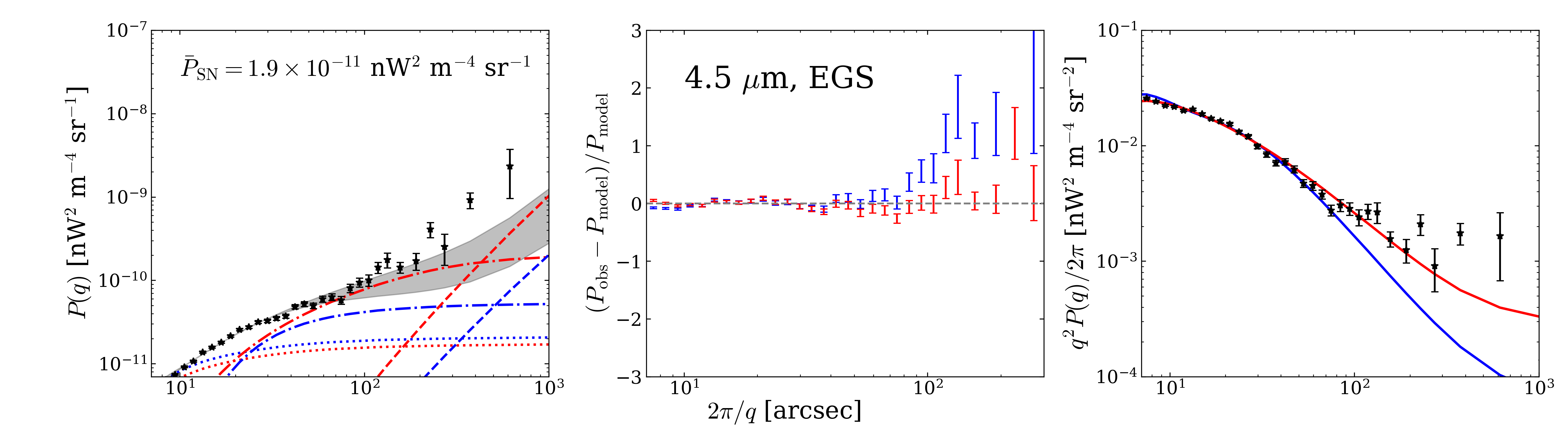}
    \caption{CIB reconstructions using the HRK12 empirical model for each SEDS angular power spectrum (represented by the black error bars). The blue lines are of the LFE limit and the red lines the HFE limit. For each FE model, the dotted, dash-dotted, and dashed lines show contributions from the shot noise, 1-halo, and 2-halo power respectively. The shaded gray area represents the allowed range between the LFE and HFE models. The left column shows the power spectrum, the middle columm shows the relative error in the model bracketing, and the right column shows the fluctuation spectrum. The average shot noise level $\bar{P}_{\rm SN}$ = mean($P_{\rm SN}^{\mathrm{LFE,HFE}}$) is shown in the left column for each power spectrum. Errors are reported at the 1$\sigma$ level.}
    \label{fig:SEDS_vertical_triple_panels}
\end{figure*}
\begin{figure*}[ht]
    \centering
    \includegraphics[width=\textwidth]{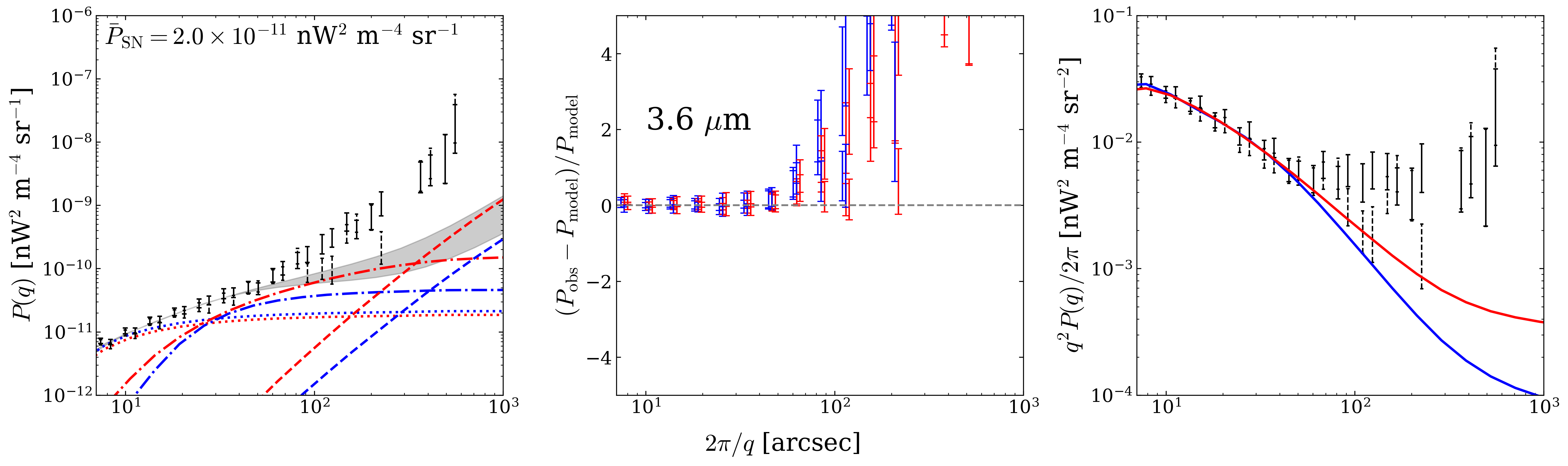}
    \includegraphics[width=\textwidth]{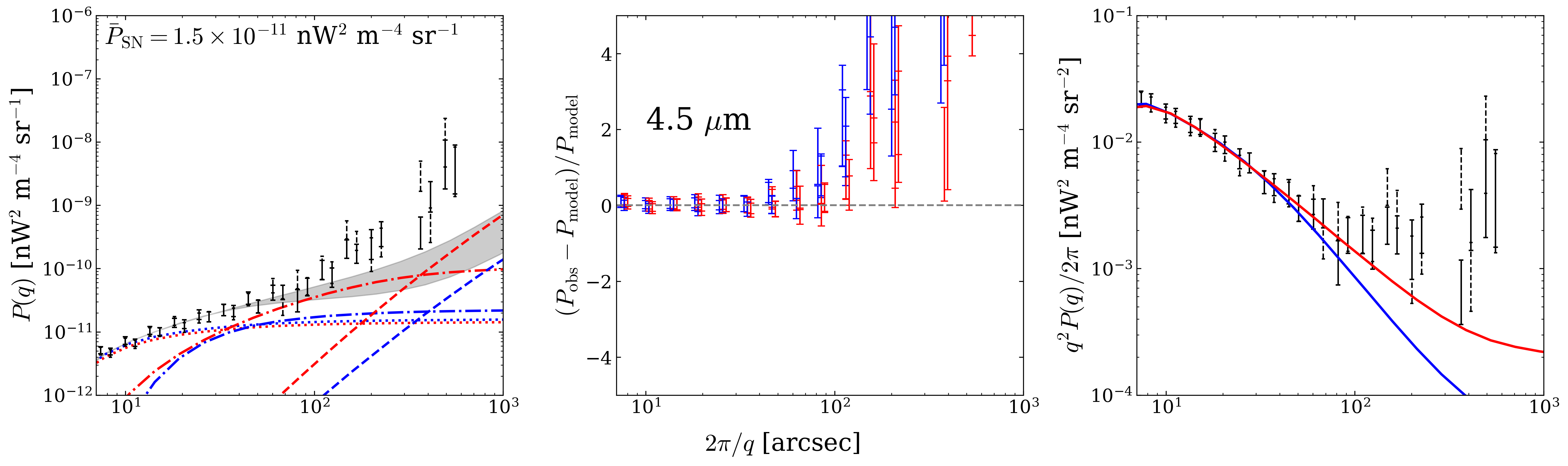}
    \caption{CIB reconstructions using the HRK12 empirical model for each GOODS angular power spectrum (represented by the black error bars). The solid and dashed error bars are of the HDF-N E1/E2 and CDF-S E1/E2 power spectra respectively. The blue lines are of the LFE limit and the red lines the HFE limit. For each FE model, the dotted, dash-dotted, and dashed lines show contributions from the shot noise, 1-halo, and 2-halo power respectively. The shaded gray area represents the allowed range between the LFE and HFE models. The left column shows the power spectrum, the middle columm shows the relative error in the model bracketing, and the right column shows the fluctuation spectrum. The average shot noise level $\bar{P}_{\rm SN}$ = mean($P_{\rm SN}^{\mathrm{LFE,HFE}}$) is shown in the left column for each power spectrum. Errors are reported at the 1$\sigma$ level.}
    \label{fig:GOODS_vertical_triple_panels}
\end{figure*}

\begin{figure*}[ht]
    \centering
    \includegraphics[width=\textwidth]{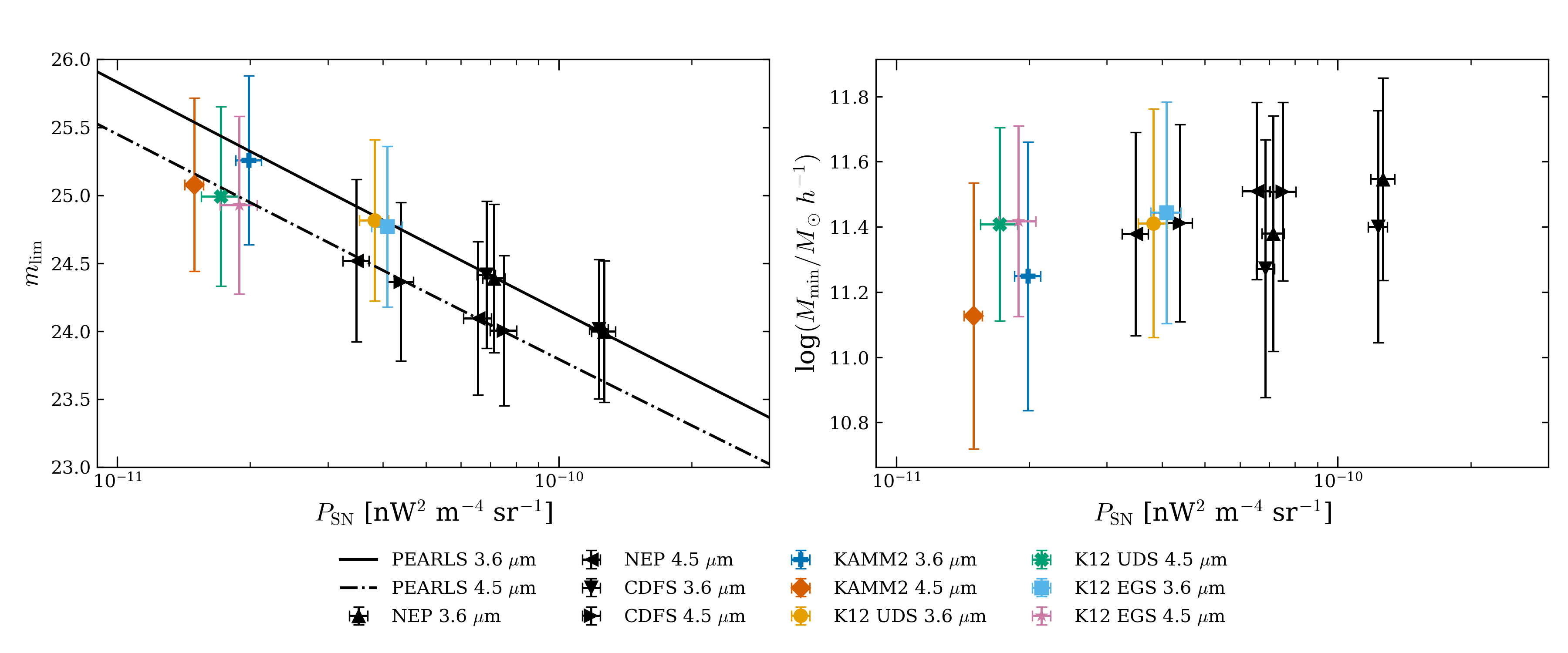}
    \caption{Bracketed parameters characterizing known galaxy populations with respect to shot noise level $P_{\rm SN}$ are shown with the same markers as in Fig. \ref{fig:cib_power_spectra_data}. In each plot error bars do not represent uncertainties but rather the limits due to the LFE/HFE spread (see Table \ref{tab:fe_summary}). The parameters shown are $m_{\rm lim}^{\rm LFE,HFE}$ (\textit{left}) and $\log(M_{\min}^{\rm LFE,HFE}/M_{\odot}h^{-1})$ (\textit{right}). The inferred $m_{\rm lim}$ values are consistent with what is expected from JWST PEARLS with the solid and dash-dotted black lines representing the $m_{\rm lim}$ values at 3.6 and 4.5 $\mu$m respectively \citep{Windhorst2023}. Across all Spitzer datasets a decrease in $\log(M_{\min}/M_{\odot} h^{-1})$ is seen which is consistent with \cite{Helgason2017}.}
    \label{fig:CIB_HOD_plot}
\end{figure*}

\begin{deluxetable*}{lllccc}
\tablecaption{Summary of Bracketed Parameters \label{tab:fe_summary}}
\tablehead{
Dataset & $\lambda$ ($\mu$m) & Field(s) & $m_{\mathrm{lim}}^{\mathrm{LFE,HFE}}$ & $\bar{P}_{\rm SN}$ & $\log M_{\min}^{\mathrm{LFE,HFE}}$ \\ 
}
\startdata
LIBRAS & 3.6 & NEP & 23.5, 24.5 & 12.7 (151.9) & 11.9, 11.2 \\
 &  &  & 23.8, 24.9 & 7.1 (85.7) & 11.7, 11.0 \\
 & 3.6 & CDFS & 23.5, 24.5 & 12.3 (148.0) & 11.8, 11.0 \\
 &  &  & 23.9, 25.0 & 6.9 (82.3) & 11.7, 10.9 \\
 & 4.5 & NEP & 23.5, 24.7 & 6.6 (98.3) & 11.8, 11.2 \\
 &  &  & 23.9, 25.1 & 3.5 (52.2) & 11.7, 11.1 \\
 & 4.5 & CDFS & 23.5, 24.6 & 7.5 (112.7) & 11.8, 11.2 \\
 &  &  & 23.8, 24.9 & 4.4 (65.8) & 11.7, 11.1 \\
SEDS & 3.6 & UDS & 24.2, 25.4 & 3.8 (45.9) & 11.8, 11.1 \\
 & 3.6 & EGS & 24.2, 25.4 & 4.1 (49.1) & 11.8, 11.1 \\
 & 4.5 & UDS & 24.3, 25.7 & 1.7 (25.7) & 11.7, 11.1 \\
 & 4.5 & EGS & 24.3, 25.6 & 1.9 (28.3) & 11.7, 11.1 \\
GOODS & 3.6 & HDF-N, CDF-S E1/E2 & 24.6, 25.9 & 2.0 (23.8) & 11.7, 10.8 \\
 & 4.5 & HDF-N, CDF-S E1/E2 & 24.4, 25.7 & 1.5 (22.4) & 11.5, 10.7 \\
\enddata
\tablecomments{$\bar{P}_{\rm SN}$ is the average of LFE/HFE values, in units of $10^{-11}$ nW$^2$ m$^{-4}$ sr$^{-1}$. The associated values in parentheses are in units of nJy $\cdot$ nW m$^{-2}$ sr$^{-1}$. $M_{\rm min}$ is in $M_{\odot}\,h^{-1}$. There are two lines for each LIBRAS wavelength and field corresponding to different iterations of the source model (and hence shot noise levels).}
\end{deluxetable*}

The bracketed parameters for each of the angular power spectra employed can be found in Table \ref{tab:fe_summary}. The reconstructed CIB power spectra for the LIBRAS, SEDS, and GOODS datasets can be found in Figs. \ref{fig:LIBRAS_vertical_triple_panels_li}, \ref{fig:LIBRAS_vertical_triple_panels_hi},
\ref{fig:SEDS_vertical_triple_panels}, and \ref{fig:GOODS_vertical_triple_panels} respectively. The left, middle, and right subplots of each figure correspond to the power spectra, relative errors, and fluctuation spectra of each dataset respectively. As in previous works we find that the CIB anisotropies at sub-arcminute scales can be attributed to faint galaxies which are unresolved and unmasked. Unsurprisingly, the relative errors $\Delta P=(P_{\rm obs}-P_{\rm model})/P_{\rm model}$ at $\lesssim0.5'$ change as a function of shot noise level and the relative contribution of the new populations becomes higher. For the LIBRAS and K12 power spectra $\Delta P<5\%$, while the KAMM4 power spectra deviate by almost $\Delta P \lesssim 10\%$ (although the Poisson errors associated with this dataset are higher than the shallower ones). At large scales ($\gtrsim 100''$), the relative errors exceed $\sim 100\%$ and grow, showing that the large-scale CIB power observed by Spitzer originates from a new population of sources that are independent of the known galaxy populations that produce the small-scale signal(s). These results also suggest that other hypothetical contributions to the source-subtracted CIB fluctuations must simultaneously contribute negligibly at sub-arcminute scales yet produce the overwhelming majority of the large-scale power. This condition may be satisfied by the proposed high-$z$ explanation \citep{Kashlinsky2018} but is generally not satisfied by the IHL one \citep{Cooray2012,Helgason2014,Thacker2015,Kashlinsky2018}.

Across all of the Spitzer datasets $\log(M_{\min}/M_{\odot}h^{-1})$ is to the order of $\sim 11$ although a mild evolution with shot noise level is observed (see Fig. \ref{fig:CIB_HOD_plot}). To quantify this we implement a modified Akaike Information Criterion (AIC; \citealt{Akaike1974}) test which includes a correction term for small sample sizes \citep{CAVANAUGH1997}. We merge our results at both 3.6 and 4.5 $\mu$m and find that the relation
\begin{equation}
\log(M_{\min}/M_{\odot}h^{-1}) = \mathcal{A}+\mathcal{B}\log(P_{\rm SN}/{\rm nW}^2\ {\rm m}^{-4}\ {\rm sr}^{-1})
\label{eq:Mmin_Psn_relation}
\end{equation}
best describes our constraints for each FE scenario, as opposed to a constant evolution, with $\mathcal{A}^{\rm LFE,\ HFE}=13.6,13.4$ and $\mathcal{B}^{\rm LFE,\ HFE}=0.18,0.23$. We also test that when dividing our results by wavelength the modified AIC still prefers a linear relationship. It is important to note that this relation is very simplistic, but nevertheless demonstrates that $\log (M_{\min}/M_{\odot}h^{-1})$ evolves with respect to $P_{\rm SN}$ (and thus $m_{\rm lim}$) at the magnitude depths we investigate. Across all of the datasets evaluated we find that $\log(M_{\min}/M_{\odot}h^{-1})\lesssim\log(M_{\rm peak}/M_{\odot}h^{-1})$, the characteristic halo mass at which the star formation rate peaks (which varies little for $z\lesssim4$) (e.g., \citealt{Behroozi2013_lackshmr}; \citealt{Ishikawa2020}; \citealt{Shuntov2022}; \citealt{Paquereau2025}; \citealt{Chaikin2025}). However, we do find that our inferred $\log (M_{\rm sat}/M_{\odot}h^{-1})$ values are much closer to (or exceed) $\log(M_{\rm peak}/M_{\odot}h^{-1})$ depending on the shot noise level. This indicates that the evolution of the 1-halo signal with shot noise probes the transition halo mass regime where star formation is at its maximum efficiency. Further narrowing the LFE/HFE bounds on our reconstructed LFs is needed in order to precisely quantify this evolution.

The constraints we place on $\log (M_{\min}/M_{\odot}h^{-1})$ are $\gtrsim 2$ orders of magnitude greater than the $\log (M_{\rm min}/M_{\odot}h^{-1})\approx8.85$ assumption made in HRK12. However, this estimate was revised in \cite{Helgason2017} to $\log (M_{\rm min}/M_{\odot}h^{-1})\approx 11.6$ after finding the large-scale clustering in HRK12 to be systematically lower than predictions from the Munich SAM \citep{Springel2005,Henriques2015} when probing galaxies at the same magnitude depths. This updated $\log (M_{\rm min}/M_{\odot}h^{-1})$ value is chosen by evaluating the halo mass distribution of galaxies contributing to the CIB fluctuations, with $11.6$ being an acceptable value across several magnitude ranges \citep{Helgason2017}. Furthermore, in selecting a more physically-motivated $\log(M_{\rm min}/M_{\odot}h^{-1})$ it is also shown that this value is sensitive to the chosen $m_{\lim}$. As stated previously, we recover this inverse relation between $\log(M_{\rm min}/M_{\odot}h^{-1})$ and $m_{\rm lim}$.

We can also test our bracketed HOD model against the bias parameter estimates of \cite{Kaminsky2025} where the angular power spectrum of faint JWST-galaxies ($m_{\rm AB} > 25$) is computed in the F444W band. Over the redshifts $0\leq z<6$ they find that $P_{\rm SN} \approx1.92\times10^{-11}$ nW$^2$ m$^{-4}$ sr$^{-1}$ and compute the average, large-scale galaxy bias in the smaller $z$-intervals $0\leq z<3$ and $3\leq z<6$. For each $z$-interval they obtain best-fit values of $1.03_{-0.10}^{+0.10}$ and $3.21_{-0.25}^{+0.24}$ respectively which correspond to $ \log (M_{\rm h}/M_{\odot}h^{-1})\sim10-11.25$ with no discernible redshift evolution due to the Poisson uncertainties in the measured angular power spectra and the width of the redshift intervals. This range is lower albeit consistent with our constraints on $\log(M_{\min}/M_{\odot}h^{-1})$ for similar shot noise values, such as the K12 bracket (specifically in the EGS field) which has $P_{\rm SN} \approx 1.9\times10^{-11}$ nW$^2$ m$^{-4}$ sr$^{-1}$ (see Table \ref{tab:fe_summary}). Specifically, we find that our $\log(M_{\min}^{\rm HFE}/M_{\odot}h^{-1})$ value is in closer agreement than the respective $\log(M_{\min}^{\rm LFE}/M_{\odot}h^{-1})$ value which makes sense given that the COSMOS-Web galaxy counts lie between the DFE and HFE models (see Fig. \ref{fig:number_counts_comparison}).

Overall there is a very good agreement between the reconstructed CIB anisotropies from known galaxies and the measurements as demonstrated in the middle panels of Figs. \ref{fig:LIBRAS_vertical_triple_panels_li}, \ref{fig:LIBRAS_vertical_triple_panels_hi}, \ref{fig:SEDS_vertical_triple_panels}, and \ref{fig:GOODS_vertical_triple_panels}. We now move to discuss the various systematic uncertainties in this reconstruction. 

\subsection{Systematic Uncertainties}\label{subsec:syst_unc}

Our reconstruction methodology does not directly account for the mask effects, which may lead to mode coupling and may be important in precisely reconstructing the various components of the source-subtracted CIB anisotropies. The effects of this were discussed in Appendix B of \cite{Kashlinsky2012}, where the individual theoretical power models were simulated in the presence of masks from the two (much smaller than the LIBRAS fields) SEDS fields from Table \ref{tab:Spitzer_fields}. As shown there, the shot noise power is not affected by the masks, the 1-halo term is affected only marginally, and a small amount of power from the new component may be transferred to sub-arcminute scales. Given the size of the two LIBRAS fields, it is impractical to run similar simulations as was done for the SEDS data. However, the consistency between our reconstructed power spectra and the observed power spectra shown throughout the middle panels in Figs.  \ref{fig:LIBRAS_vertical_triple_panels_li}, \ref{fig:LIBRAS_vertical_triple_panels_hi},
\ref{fig:SEDS_vertical_triple_panels},
and \ref{fig:GOODS_vertical_triple_panels} empirically demonstrate that the masking effects are small and at most potentially lead to some small power transfer to scales $\lesssim10''$.
Indeed we find that different fields, and thus different mask geometries, at comparable shot noise levels produce consistent $m_{\rm lim}^{\rm LFE,\,HFE}$ and  $\log(M_{\min}^{\rm LFE,\, HFE}/M_{\odot}h^{-1})$ constraints. This strongly indicates that the masking effects across all of the Spitzer fields analyzed in this work are negligible. Furthermore, we also quantify residual calibration errors in the LIBRAS data. In Appendix \ref{sec:cal_syst} we find that such errors are unlikely to contribute to the LIBRAS angular power spectra.
\begin{figure}
    \centering
    \includegraphics[width=0.7\columnwidth]{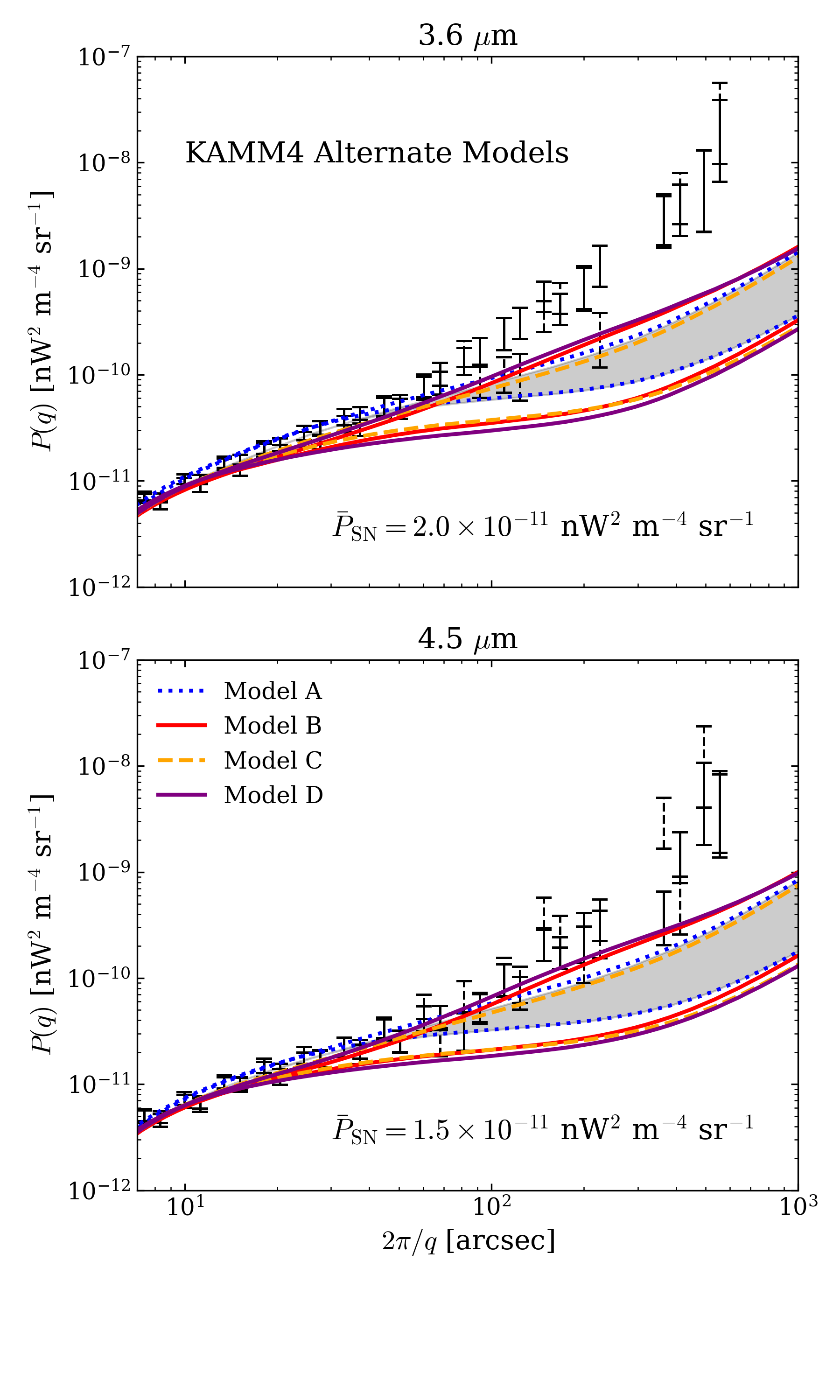}
    \caption{Alternate power spectrum models for the KAMM4 3.6 $\mu$m ({\it top}) and 4.5 $\mu$m ({\it bottom}) datasets. Models A (dotted blue line), B (solid red line), C (dashed yellow), and D (solid purple line) are characterized by changing the base HMF and concentration-mass relation to those from \cite{Ishiyama2015} and \cite{Ludlow2016} respectively, substituting the $\alpha_{\rm sat}$ and $\Delta_{\rm sat}$ values both to 1, allowing $\log(M_{\min}/M_{\odot}h^{-1})$ to evolve with redshift, and finally combining the previous three alterations respectively. The base model is shaded in gray.}
    \label{fig:KAMM07_alt_models}
\end{figure}

It is also important to test how our CIB reconstructions are affected by the assumptions made in our halo modeling. Hereafter we refer to the model described in \S \ref{subsec:methods} -- \ref{subsec:results} as the base model and construct four different alternate models:

\begin{enumerate}
    \item (Model A) We swap the HMF and concentration-mass relation to those from \cite{Ishiyama2015} and \cite{Ludlow2016} respectively while retaining the constrained parameters provided in Table \ref{tab:fe_summary}.
    \item (Model B) We change the $\alpha_{\rm sat}$ and $\Delta_{\rm sat}$ values both to 1 while retaining the constrained parameters provided in Table \ref{tab:fe_summary}.
    \item (Model C) We impose a mild redshift evolution in $\log(M_{\min}/M_{\odot}h^{-1})$ motivated by \cite{Contreras2017}. We assume the function \begin{equation}
    \log(M_{\min}^{\rm LFE,\, HFE}/M_{\odot}h^{-1}) =
    \begin{cases}
    \log(M_{\min,0}^{\rm LFE,\, HFE}/M_{\odot}h^{-1}), & \text{if } z < 1, \\
    \log(M_{\min,0}^{\rm LFE,\, HFE}/M_{\odot}h^{-1})+\gamma_{M_{\min}}(z-1),  & \text{if } z \ge 1
    \end{cases}
    \end{equation}
    where we take $\log(M_{\min,0}^{\rm LFE,\, HFE}/M_{\odot}h^{-1})$ to be the parameters constrained from the base model (see Table \ref{tab:fe_summary}) and assume $\gamma_{M_{\rm min}}=-0.10$.
    \item (Model D) We take each of the changes presented in Models A, B, and C and evaluate the angular power spectra using the constrained parameters provided in Table \ref{tab:fe_summary}.
\end{enumerate}
We implement these four models across the whole suite of Spitzer datasets, although for conciseness we only present those for KAMM4 (see Fig. \ref{fig:KAMM07_alt_models}). We find that changing the $\alpha_{\rm sat}$ and $\Delta_{\rm sat}$ parameters and evolving $\log(M_{\min}/M_{\odot}h^{-1})$ as a function of redshift (corresponding to Models B and C respectively) decreases the amplitude of the 1-halo power while changing the HMF and concentration-mass relation (Model A) increases the 1-halo power slightly at smaller scales (see Fig. \ref{fig:KAMM07_alt_models}). At scales $\lesssim0.5'$ the 1-halo power produced by the alternate and base models differ on average between $\sim10-30\%$. Such tests show that even by varying the assumptions used in our bracketing methodology we find galaxies hosted in DM halos of $\log (M_{\rm min}/M_{\odot}h^{-1})\gtrsim11$ produce the clear majority of the small-scale power and cannot account for the large-scale power measured across the suite of Spitzer datasets utilized in this work. The systematic uncertainties may dominate over the cosmic variance of the CIB anisotropic signal, but are still small overall.

\section{Reconstructed CIB anisotropies from known galaxies in  ongoing space surveys}

In this section we discuss the prospects of probing the source-subtracted CIB anisotropies from the new populations in light of the reconstructed contributions from remaining known galaxies obtained in this study. Given that these known galaxy populations will dominate other additional low-$z$ components (e.g. IHL), we can extrapolate the discussion of the contributions from remaining known galaxies to shorter wavelengths when probing the origin of the new component there. This probe is potentially possible with the current space missions conducting near-IR sky surveys over large regions: Euclid, Roman, and SPHEREx (for possible use of JWST for such measurement see \citealt{Kashlinsky2015b}). 

We emphasize again that the Spitzer source-subtracted CIB fluctuations from new populations appear at 3.6, 4.5 $\mu$m after removing identified sources to $m_{\rm AB}\gtrsim 25$ and reaching shot noise levels (at 3.6, 4.5 $\mu$m) below $P_{\rm SN}\lesssim(24,\ 22)$ nJy$\cdot$nW m$^{-2}$ sr$^{-1}$ (which are limits within the scope of the Euclid and Roman surveys). 
In what follows we illustrate the feasibility of probing the CIB from the new populations discovered in Spitzer and later AKARI data, which directly cover $(2-4.5)$ $\mu$m. 

We show that Roman has by far the best prospects for this measurement given its depth, wavelengths, resolution, planned surveys, and negligible systematics of relevance. Euclid closely follows next with both its Wide and Deep Surveys. We demonstrate that SPHEREx cannot be used here, given its depth and resolution where the CIB fluctuations from the remaining known galaxies would clearly dominate the source-subtracted CIB detected in the Spitzer data. 

%{\bf Revise some phrasings below. Interchange figs 10 and 11.}

\subsection{General requirements for probing the source-subtracted CIB anisotropies}
\begin{figure*}[ht!]
    \centering
    \includegraphics[width=\textwidth]{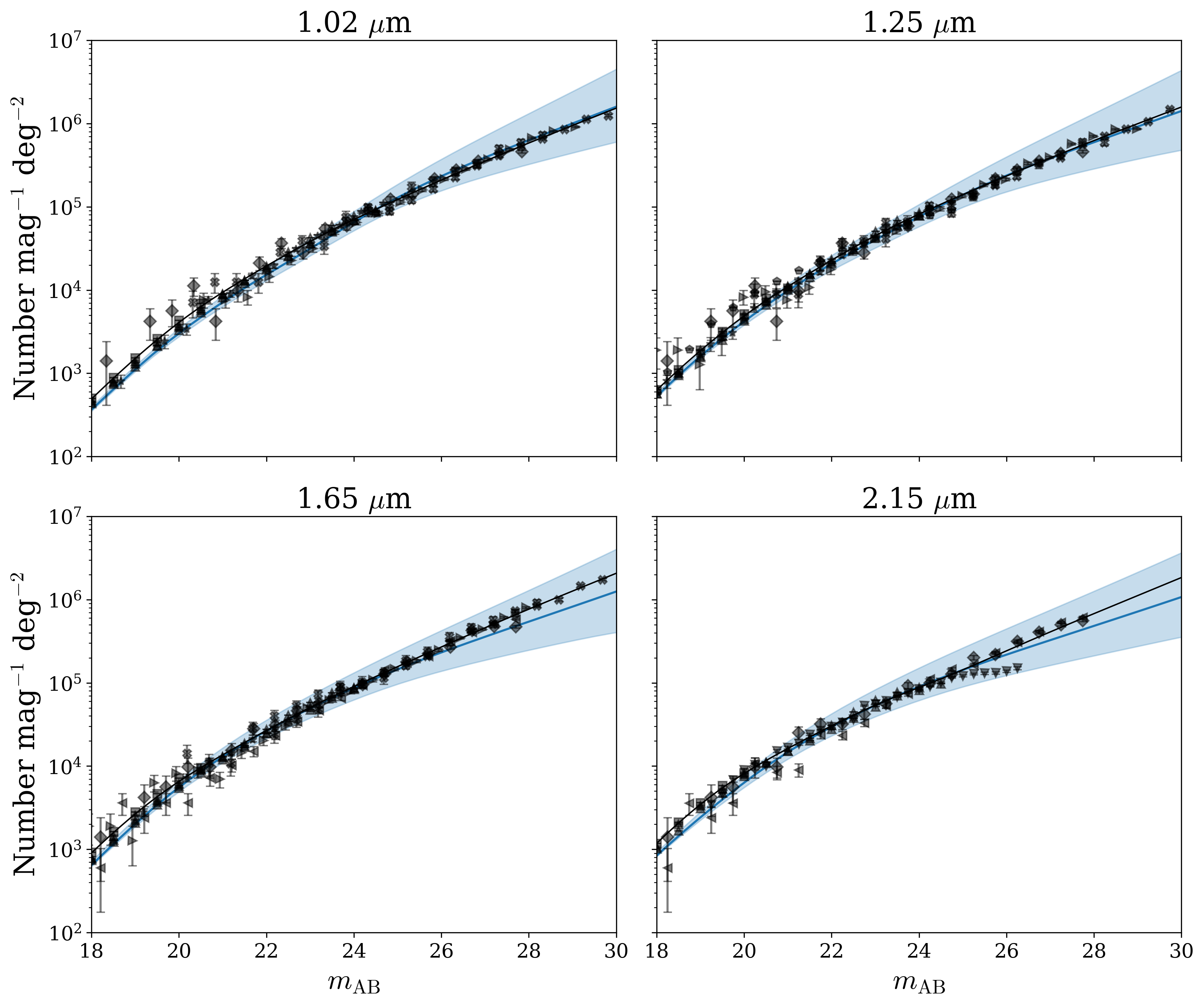}
    \caption{Number counts are shown in several NIR wavelengths. These are derived by interpolating over the original HRK12 wavelengths, with the shaded blue region bracketing the LFE and HFE limits and the solid blue line marking the DFE model. A multitude of galaxy count data (for more detail we refer the reader to \cite{Windhorst2023}) is shown to be consistent with the empirical HRK12 reconstruction. Solid black lines show the interpolated counts fits from the PEARLS JWST survey \citep{Windhorst2023}.}
    \label{fig:HRK_future}
\end{figure*}

The purpose of future measurements with instruments, depths and wavebands different from Spitzer would be to probe and identify the nature of the Spitzer-found CIB anisotropy signal. Thus, the new reconstruction should be used when deciding on the feasibility of science undertakings designed to probe the source-subtracted CIB fluctuations at higher accuracy, identify directly their epochs from the Lyman-break feature as one moves to shorter $\lambda$, and isolate the shot noise power produced by new sources, as well as to 
explore the nature of the new sources, being black hole and/or stellar populations, from the accurate measurement of the CXB-CIB cross-power \citep{Kashlinsky2019}.

\begin{figure*}[ht!]
    \centering
    \includegraphics[width=\textwidth]{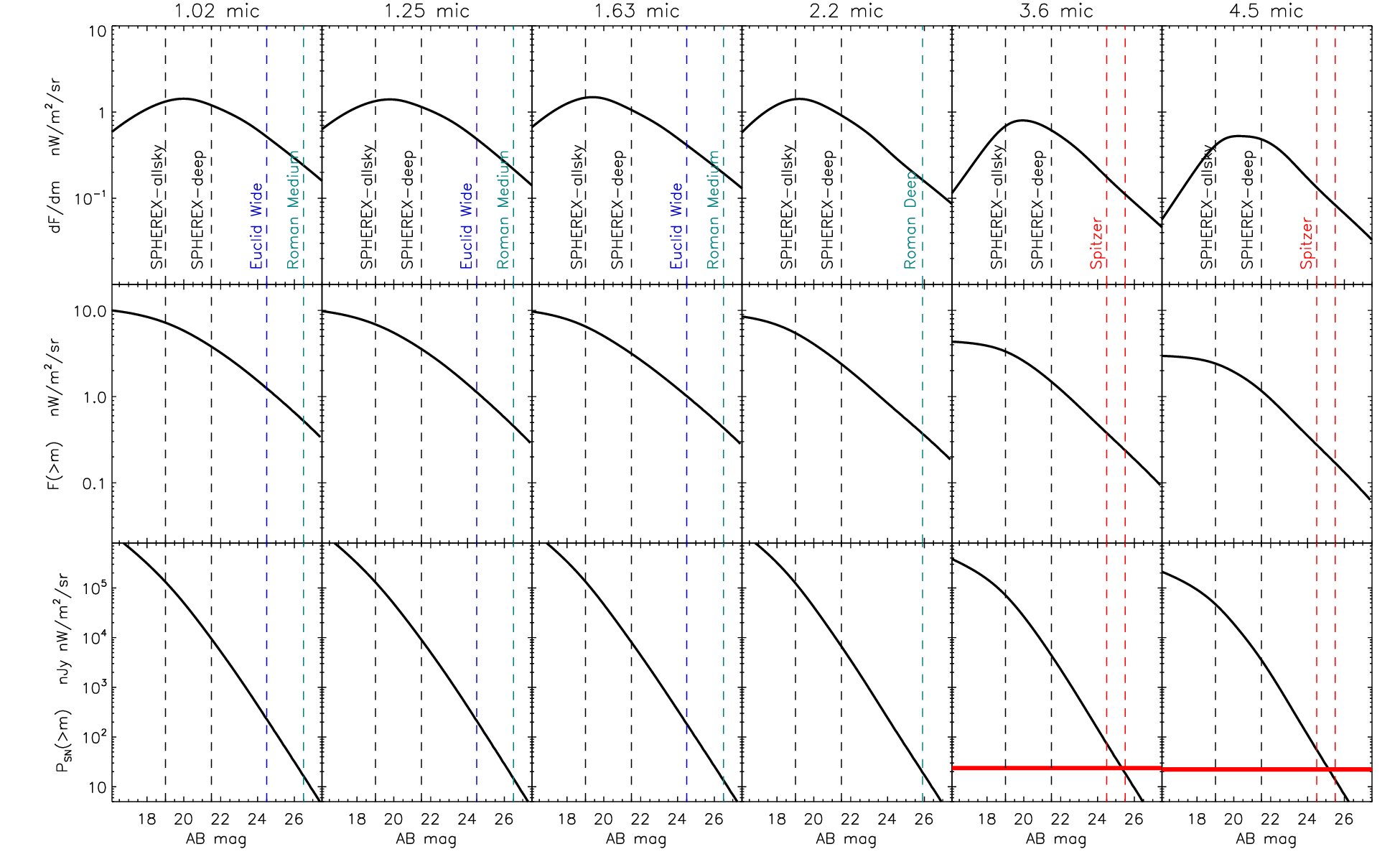}
    \caption{The rate of accumulation of the CIB flux from known galaxies $dF/dm$ is shown at wavelengths overlapping with the surveys in Table \ref{tab:forthcoming} is shown in the upper row. Middle row shows the next CIB flux from the remaining known populations, $F(>m)$. The bottom row shows the shot noise power in the remaining known populations, $P_{\rm SN}(>m)$ as function of the survey magnitude depth $m$. The thick solid lines correspond to the interpolated number counts from the PEARLS JWST survey \citep{Windhorst2023}. Vertical dashes show the parameter of the prospective surveys: SPHEREx in black, Euclid in blue, Roman in green. Red dashes mark the parameters of the Spitzer measurements from Table \ref{tab:fe_summary}. }
    \label{fig:future}
\end{figure*}

We start here by demonstrating the accuracy of the reconstruction when extrapolating to the near-IR wavelengths shorter than used in the Spitzer measurements discussed earlier. After demonstrating the good fits to the galaxy counts at the shorter wavelengths \citep{Windhorst2023}, the contribution of known galaxy populations to the source-subtracted CIB fluctuations are calculated as a function of magnitude over a suite of NIR wavelengths covered by the ongoing space missions. 
%In order to implement our  reconstructions of the CIB anisotropies from known galaxy populations, we interpolate between the observer-frame LFs in the original fiducial bands presented in HRK12. 

As seen in Fig. \ref{fig:HRK_future}, these empirical reconstructions are highly consistent with those from JWST PEARLS \citep{Windhorst2023}. Additionally, extrapolated galaxy counts from \cite{Tompkins2025} at $1.54$ $\mu$m are within the FE limits presented in Fig. \ref{fig:HRK_future} (see their Fig. 25), further confirming the reconstruction methodology of HRK12. The subsequent angular power spectra at these new wavelengths are then computed using the base model introduced in \S \ref{subsec:methods}. We use the best-fit parameters satisfying Eq. \ref{eq:Mmin_Psn_relation} for each FE limit and extrapolate to the shot noise levels probed across all of the Roman, Euclid, and SPHEREx surveys. The fluctuation spectra from remaining known galaxies measured by these missions are presented in the figures below for each of the 3 missions.

\begin{deluxetable*}{clcccccc}
\tablecaption{Summary of forthcoming near-IR datasets \label{tab:forthcoming}}
%\tablehead{\hspace{-2mm}\colhead{Dataset} & 
\tablehead{\colhead{Dataset} & 
\colhead{Survey} &
\colhead{Pixel} &
\colhead{Area} & 
\colhead{Bands} & 
\colhead{Point Source Depth} &
\colhead{$F(>m)$} &
\colhead{$P_{\rm SN}(>m)$}\\
\colhead{ } & 
\colhead{ } &
\colhead{ } &
\colhead{ } & 
\colhead{ } & 
\colhead{5$\sigma$ $m_{\rm AB}$} &
\colhead{nW m$^{-2}$ sr$^{-1}$} &
\colhead{nJy $\cdot$ nW m$^{-2}$ sr$^{-1}$}
}
\startdata
Euclid  & Wide & $0.3''$ & 14,679 deg$^2$  & $Y_{\rm E}$, $J_{\rm E}$, $H_{\rm E}$ & 24.4, 24.4, 24.4  & 1.4, 1.2, 0.9 & 277, 242, 194 \\
        & Deep & $0.3''$ & 53 deg$^2$ & $Y_{\rm E}$, $J_{\rm E}$, $H_{\rm E}$ & 26.4, 26.4, 26.4  & 0.6, 0.5, 0.3 & 21, 17, 12 \\
% Euclid  & Wide & $0.3''$ & 14,679 deg$^2$  & $Y_{\rm E}$, $J_{\rm E}$, $H_{\rm E}$ & 24.4, 24.4, 24.4  & 1.44, 1.18, 0.89 & 277, 242, 194 \\
 %       & Deep & $0.3''$ & 53 deg$^2$ & $Y_{\rm E}$, $J_{\rm E}$, $H_{\rm E}$ & 26.4, 26.4, 26.4  & 0.63, 0.48, 0.34 & 20.9, 16.6, 12.0 \\
\hline
Roman   & Wide & $0.1''$ & 5,117 deg$^2$ & $H$ & 26.2 & 0.4 & 16 \\
  & Medium & $0.1''$ & 2,415 deg$^2$ & $Y$, $J$, $H$ & 26.5, 26.4, 26.4  & 0.6, 0.5, 0.3 & 18, 17, 12 \\
 & Deep &  $0.1''$ & 19.2 deg$^2$ & $z, Y, J, H, F, K$ & 27.7$\rightarrow$25.9  & 0.5$\rightarrow$0.3 & 4.5$\rightarrow$16.5 \\
 & Ultra Deep & $0.1''$ & 5 deg$^2$ & $Y$, $J$, $H$ & 28.2, 28.2, 28.1  & 0.3, 0.2, 0.1 & 1.8, 1.3, 1.1 \\
 & HLTDS-$w$ & $0.1''$ & 18.27 deg$^2$ & $R$, $z$, $Y$, $J$, $H$ & 28.8 $\rightarrow$ 28.9  & 0.4 $\rightarrow$ 0.1 & 2.6 $\rightarrow$ 0.3 \\
 & HLTDS-$d$ & $0.1''$ & 6.47 deg$^2$ & $z$, $Y$, $J$, $H$, $F$ & 29.4 $\rightarrow$ 29.3  & 0.2 $\rightarrow$ 0.1 & 0.4 $\rightarrow$ 0.2 \\
%    & Deep &  $0.1''$ & 19.2 deg$^2$ & $z$, $Y$, $J$, $H$, $H/K$, $K_{\rm s}$ & 27.7, 27.7, 27.6, 27.5, 27.0, 25.9 \\
% Roman   & Wide & $0.1''$ & 2,702 deg$^2$ & $H$ & 26.2 & 0.37 & 16.0 \\
%   & Medium & $0.1''$ & 2,415 deg$^2$ & $Y$, $J$, $H$ & 26.5, 26.4, 26.4  & 0.60, 0.48, 0.34 & 18.3, 16.6, 12.0 \\
%  & Deep &  $0.1''$ & 19.2 deg$^2$ & $Z, Y, J, H, F, K$ & 27.7$\rightarrow$25.9  & 0.46$\rightarrow$0.28 & 4.46$\rightarrow$16.15 \\
%  & Ultra Deep & $0.1''$ & 5 deg$^2$ & $Y$, $J$, $H$ & 28.2, 28.2, 28.1  & 0.28, 0.20, 0.14 & 1.78, 1.33, 1.07 \\
% %    & Deep &  $0.1''$ & 19.2 deg$^2$ & $z$, $Y$, $J$, $H$, $H/K$, $K_{\rm s}$ & 27.7, 27.7, 27.6, 27.5, 27.0, 25.9 \\
\hline
% SPHEREx & All Sky & $6.15''$ & 4$\pi$ sr & 0.75-3.8, 3.8-5.0 $\mu$m%\tablenotemark{a} 
% &  18.2-19.7, 17.3-18.7\\%\tablenotemark{b}\\
%    & Deep & $6.15''$ & 200 deg$^2$ & 0.75-3.8, 3.8-5.0 $\mu$m  & 21-21.5, 19.1-20.5\\%\tablenotemark{c} \\
% SPHEREx & All Sky & $6.15''$ & 4$\pi$ sr & 0.75$\rightarrow$3.8  $\mu$m%\tablenotemark{a} 
% &  18.2$\rightarrow$19.7  & 5.81$\rightarrow$3.14 & 67915$\rightarrow$45656 \\%\tablenotemark{b}\\
%    & Deep & $6.15''$ & 200 deg$^2$ & 0.75$\rightarrow$3.8 $\mu$m  & 21$\rightarrow$21.5  & 3.60$\rightarrow$1.36 & 6917$\rightarrow$3316 \\
SPHEREx & All Sky & $6.15''$ & 4$\pi$ sr & 0.75$\rightarrow$3.8  $\mu$m%\tablenotemark{a} 
&  18.2$\rightarrow$19.7  & 5.8$\rightarrow$3.1 & 67913$\rightarrow$45655 \\%\tablenotemark{b}\\
   & Deep & $6.15''$ & 200 deg$^2$ & 0.75$\rightarrow$3.8 $\mu$m  & 21$\rightarrow$21.5  & 3.6$\rightarrow$1.4 & 6917$\rightarrow$3316 \\
   %\tablenotemark{c} \\
\enddata
\tablecomments{Parameters of each forthcoming dataset are shown. The $F(>m)$ and $P_{\rm SN}(>m)$ quantities from known galaxy population are computed using the DFE model of the reconstructed galaxy counts (see Fig. \ref{fig:HRK_future}). The ``-$w$" and ``-$d$" attached to the High-Latitude Time Domain Survey (HLTDS) parameters correspond to the Wide and Deep tiers respectively. For SPHEREx the 3.8-5.0 $\mu$m range is not shown because its sensitivity is not sufficient for this science.}
\end{deluxetable*}

The signal at 3.6, 4.5 $\mu$m appears at $P_{\rm SN}\lesssim 22-24$ nJy$\cdot$nW m$^{-2}$ sr$^{-1}$, and requires populations producing the near-IR CIB flux of $F\sim 1$ nW m$^{-2}$ sr$^{-1}$. The general implications of this were presented in \cite{Kashlinsky2007b} and are as follows: the shot noise power from the new populations with the typical flux $S$ is linked to the net flux levels via $P_{\rm SN}\simeq SF$ implying very faint CIB sources of $m_{\rm AB}>29-30$. Additionally, as suggested by \cite{Matsumoto2011} the source-subtracted CIB follows approximately the Rayleigh-Jeans energy spectrum, $I_\nu\propto \lambda^{-3}$ out to at least $\lambda\simeq 2\ \mu$m. These source-subtracted CIB fluctuations appear to be highly coherent with the unresolved soft ([0.5--2] keV) CXB implying a much higher fraction of black holes among the new sources than in the known populations. Hence, to be successful the surveys must be able to (1) reach sufficiently low levels of shot noise from the remaining known galaxies, and (2) leave known galaxies at sufficiently low CIB fluxes (effectively reaching sufficiently faint magnitudes). 

The contribution of known galaxy populations to the source-subtracted CIB fluctuations are calculated over parameters suitable for the ongoing space missions, Euclid, SPHEREx and Roman. In order to implement our  reconstructions of the CIB anisotropies from known galaxy populations, we interpolate between the observer-frame LFs in the original fiducial bands presented in HRK12. The interpolated quantities of interest and relevance here (obtained using the DFE model) are provided in Table \ref{tab:forthcoming}.

To compare with the expected signal from the new populations implied by the Spitzer data at 3.6, 4.5 $\mu$m we use a template corresponding to the $\Lambda$CDM standard model projected to $z=10$ which is normalized to 3.6 $\mu$m and assumes an amplitude that scales as a Rayleigh-Jeans energy spectrum ($I_{\nu}\propto \lambda^{-3}$) to $\simeq 2$ $\mu$m as proposed by \cite{Matsumoto2011}. The CIB may not increase according to this energy spectrum or be altogether cut off by Lyman-break at the shorter wavelengths probed by these surveys, so when the longest wavelength of the survey is below $2$ $\mu$m as in the case of Euclid we also show the magnitude of the signal at 3.6 $\mu$m for reference of easy rescaling. Its characteristic shape is then shown with pink lines in the figures in this section: the shape has a characteristic (shallow) peak at the co-moving scale of the horizon scale of the matter-radiation equality projected to the selected $z$. Of course, this is an idealized template as emissions over a range of $z$ are certain to contribute to the 2D CIB power spectrum via the Limber equation as discussed in \cite{Kashlinsky2005review}, although the 3D $\Lambda$CDM templates at $z\gg1$ are sufficiently similar differing mainly by the peak angular scale which reflects the angular scale subtended by the horizon at the matter-radiation equality epoch projected to the selected $z$ (see Fig. 12 of K12 for other single $z$ templates). At the larger scales  the power spectrum of the populations traces the transition into the Harrison-Zeldovich regime, $P\propto q^{-1}$. The Harrison-Zeldovich regime should be identifiable until cirrus intervenes as shown with the gray lines in the figures below. The Lyman-tomography method proposed by \cite{Kashlinsky2015} could isolate the contributions from different $z$ for some Euclid and Roman configurations discussed below.

\begin{figure*}[ht]
    \centering
    \includegraphics[width=\textwidth]{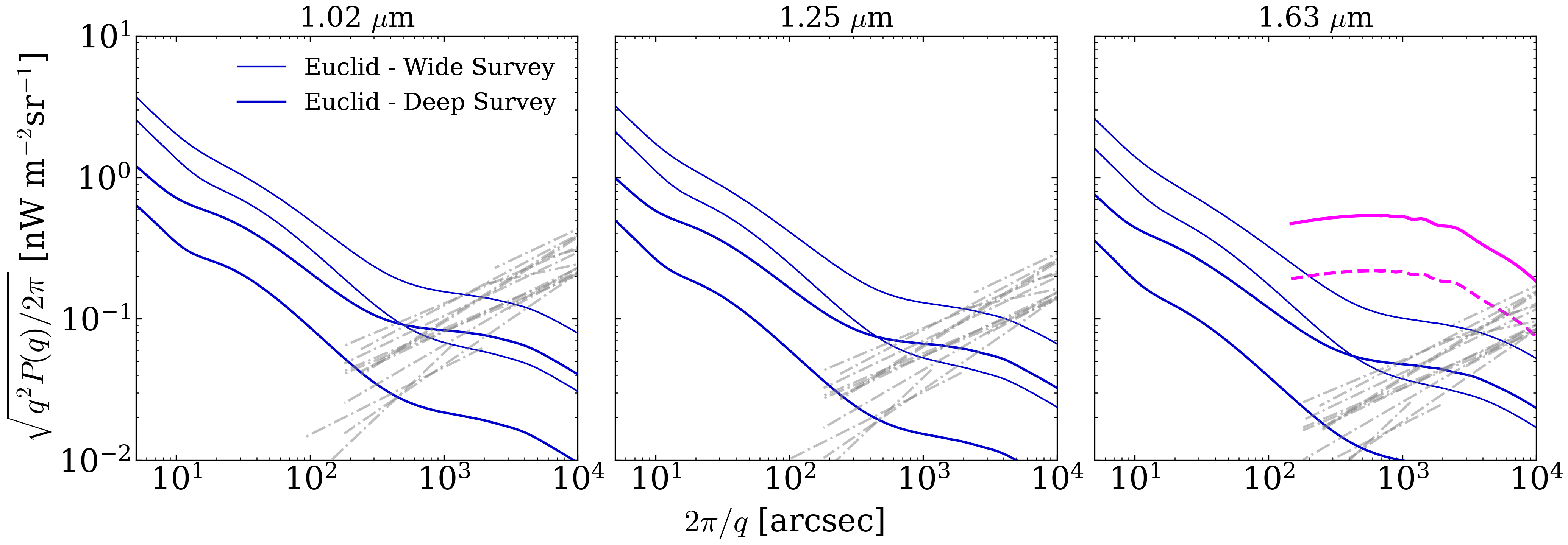}
    \caption{Predicted Euclid-observed fluctuation spectra due to known galaxy populations are shown at 1.02 ({\it left}), 1.25 ({\it middle}), and 1.63 ({\it right}) $\mu$m. The Wide Survey is shown with the solid, thin blue lines while the Deep Survey is shown in the solid, thick blue lines. The dash-dotted gray lines are estimated contributions from the ISM (see Sec. 3 of \cite{Kashlinsky2019}). The large-scale power seen in the 1.63 $\mu$m plot is of two estimates of the hypothesized high-$z$ clustering signal  where (1) the solid pink line assumes a $\Lambda$CDM model with an amplitude that scales with a Rayleigh-Jeans spectrum (normalized to the amplitude of the Spitzer 3.6 $\mu$m fluctuations) and (2) the dashed pink line shows the same $\Lambda$CDM model but with an amplitude estimated at 2.2 $\mu$m.}
    \label{fig:Euclid_fluctuation_spectra}
\end{figure*}

Foregrounds such as the ISM (or Galactic cirrus) must be considered when probing the source-subtracted CIB fluctuations with new space missions. Zodiacal light appears sufficiently smooth \citep{Arendt2016}, while the ISM power spectrum is steep and is typically modeled as $P_{\rm ISM}\propto q^{-n}$ where $n \gtrsim2$ (e.g., \citealt{Wright1998}; \citealt{Miville2002};
\citealt{Kiss2003}; \citealt{Lagache2007})
. The dot-dashed gray lines in Figs. \ref{fig:Euclid_fluctuation_spectra}-\ref{fig:SPHEREx_fluctuation_spectra} indicate various published measurements of the ISM power spectra which are compiled and translated to different wavelengths as described in \cite{Kashlinsky2019}. While there are large uncertainties due to intrinsic variation in structure and color across the sky, the steep rise of the ISM power spectrum indicates that it should exceed the high-$z$ component on scales $\gtrsim {10^4}''$. 

Given the many orders of magnitude spanned for the recovered CIB parameters from the remaining known galaxies, in the plots in this section we present the reconstructed results in terms of their $\sqrt{q^2P(q)/(2\pi)}$. We also do not convolve the powers with the beam, which is significant for SPHEREx ($\sim 10''$), but is below the shown range of angular scales for Euclid and Roman. The lines for the contributions to the CIB anisotropies by the remaining known galaxies in these surveys in the figures here follow the color notation in Fig. \ref{fig:future}: Euclid is shown with blue lines, Roman with green, and SPHEREx with black ones.

\subsection{Euclid}

The Euclid ESA mission launched in 2023 \citep{Euclid2025} presents an excellent opportunity to probe the origin of the source-subtracted CIB found in Spitzer measurements \citep{Kashlinsky2018}. Euclid has 4 photometric bands VIS in the visible covering $[0.6-0.9]\ \mu$m and the three photometric bands of the NISP instrument, $Y_E, J_E, H_E$ centered close to the standard $Y$, $J$, $H$ bands. Euclid has excellent (sub-arcsecond) angular resolution. The Wide Survey will cover about 15,000 deg$^2$ out to the limits in Table \ref{tab:forthcoming} and about 100 deg$^2$ in its Deep Survey going 2 magnitudes deeper. In addition to other CIB science discussed below, this will enable precision measurements of the source-subtracted CIB fluctuations and determination of the epochs where it was produced by cross-correlating with the $VIS$ (and $Y$) channel to probe its Lyman-break as discussed in detail in \cite{Kashlinsky2018}. This will be done within the NASA LIBRAE (Looking at Infrared Background Radiation with Euclid) approved program - \url{https://euclid.caltech.edu/page/kashlinsky-team}. Below we present the reconstructed CIB from the known galaxy populations remaining there in the Wide and Deep Surveys and the various science prospects this will impact. The reconstructions done in this work based on the recent Spitzer measurements of \cite{Kashlinsky2025} are consistent with, but significantly more accurate than, the earlier ones shown in Fig. 34 of \cite{Kashlinsky2018}.

% Euclid - cite \cite{Euclid2025}; Euclid-CIB cite \cite{Kashlinsky2018}.

\subsubsection{Wide Survey}

The Wide Survey with Euclid is going to last for 6 years and will cover the net area of 15,000 deg$^2$ largely excluding the Galactic Plane. The NISP exposures over each FOV of $0.7^\circ$ on the side with $0.3''$ pixels will reach $m_{\rm lim}\simeq 24$ in the three near-IR bands. 
Applying the reconstruction machinery developed here, Fig. \ref{fig:Euclid_fluctuation_spectra} shows the source-subtracted CIB fluctuations from galaxies remaining at fainter magnitudes than the Wide Survey limits. They are at a sufficiently low level to enable probing the CIB fluctuations from new populations as shown in pink lines. With this more accurate reconstruction and the larger Wide Survey area one would be able to probe much better the nature and epochs of the new populations behind the source-subtracted CIB. In particular, the cross-power could be probed with Spitzer in the same areas as analyzed in \cite{Kashlinsky2007data,Kashlinsky2012,Kashlinsky2025}. Additionally, the large net sky area and sufficiently deep exposures will allow us to better apply the Lyman-tomography methods proposed to identify the history of emissions and Baryonic-acoustic Oscillations at $z>10$ \citep{Kashlinsky2015}.

\subsubsection{Deep Survey}

The Euclid Deep Survey will go 2 magnitudes deeper, but cover smaller areas over parts of the sky collecting about 100 deg$^2$ of data. The thick solid lines in Fig. \ref{fig:Euclid_fluctuation_spectra} show the CIB fluctuations from known populations remaining in the Euclid Deep Survey data. The uncertainties between the LFE and HFE models are significantly smaller than in Fig. 34 of \cite{Kashlinsky2018} where the original HRK12 was used. This should allow for the robust measurement of the populations behind the Spitzer source-subtracted CIB. The shot noise remaining from known galaxy populations at the Euclid bands spanning 1--1.7 $\mu$m in the Deep Survey is only $P_{\rm SN} \simeq 12-21$ nJy$\cdot$nW m$^{-2}$ sr$^{-1}$. Doing the source-subtracted CIB measurement with better statistical accuracy achieved from the much larger sky area and at lower shot noise levels than Spitzer-based analyses may identify the shot noise power where the large-scale clustering begins to decrease, or drops, thereby potentially isolating the magnitude range of the new populations explicitly \citep{Kashlinsky2007b}.

\subsection{Roman}

\begin{figure*}[ht]
    \centering
    \includegraphics[width=\textwidth]{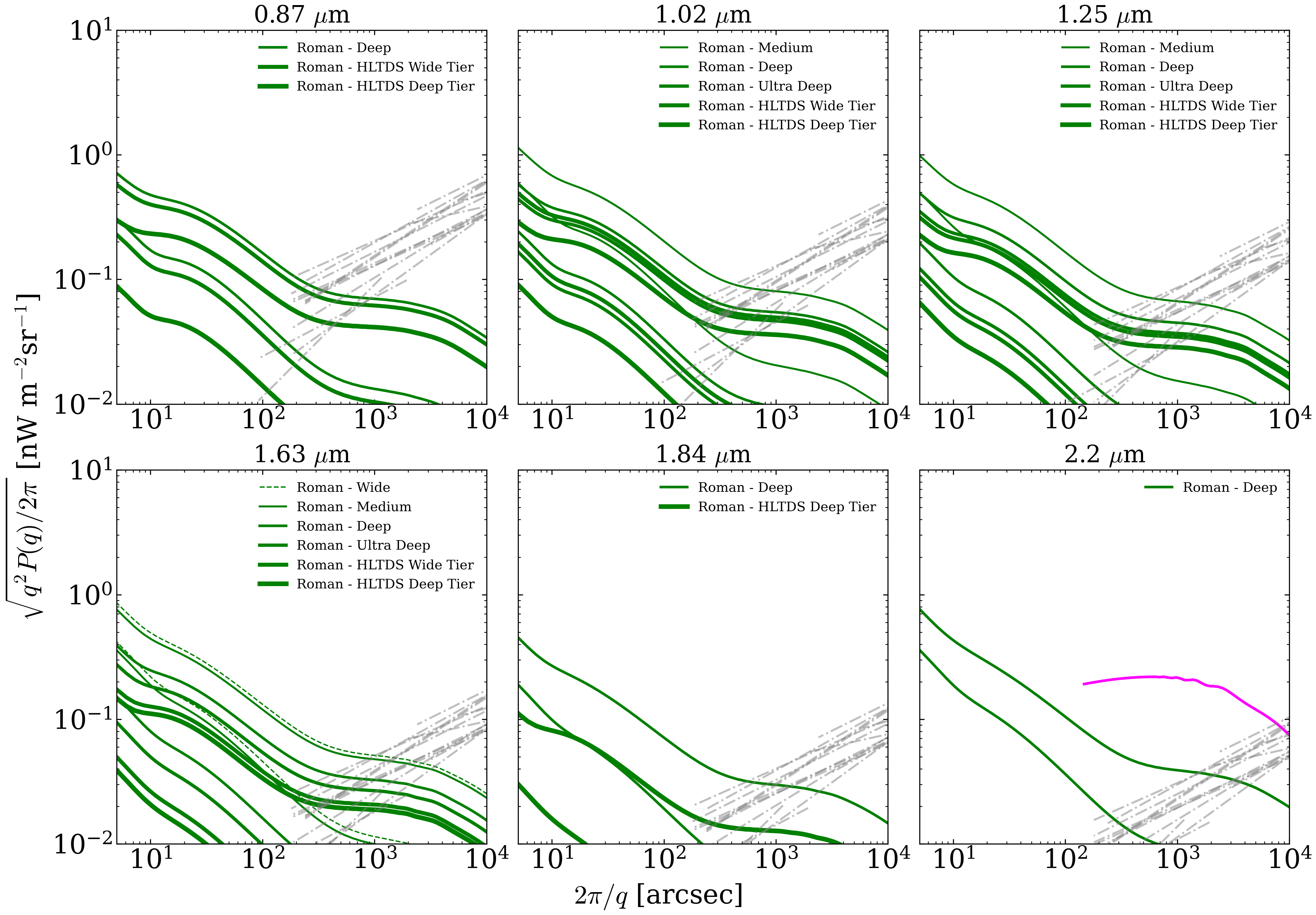}
    \caption{Predicted Roman-observed fluctuation spectra due to known galaxy populations are shown at 0.87 ({\it top left}), 1.02 ({\it top middle}), 1.25 ({\it top right}), 1.63 ({\it bottom left}), 1.84 ({\it bottom middle}), and 2.2 ({\it bottom right}) $\mu$m. The green dashed lines show the LFE and HFE reconstructions for the Wide Survey parameters. Medium Survey is shown with the solid, thin green lines, the Deep Survey is shown in the solid, medium-thick green lines, and the Ultra-Deep Survey is shown in the solid, thick green lines. The HLTDS reconstructions are also shown as well for both the Wide and Deep tiers. The dash-dotted gray lines are estimated contributions from the ISM (see Sec. 3 of \cite{Kashlinsky2019}). The large-scale power seen in the 2.2 $\mu$m plot is an estimate of the hypothesized high-$z$ clustering signal  where the solid pink line assumes a $\Lambda$CDM model with an amplitude that scales with a Rayleigh-Jeans spectrum (normalized to the amplitude of the Spitzer 3.6 $\mu$m fluctuations).}
    \label{fig:Roman_fluctuation_spectra}
\end{figure*}
The soon to be launched NASA Roman mission is going to observe large areas of the sky in numerous near-IR and visible bands with high sensitivity and $0.1''$ resolution. It plans numerous surveys  \citep{Roman2024,Roman2025} which will be of great use in identifying the origin of and the populations behind the source-subtracted CIB anisotropies. Fig. \ref{fig:Roman_fluctuation_spectra} sums up the reconstructed CIB from the remaining known galaxies in the various planned Roman surveys discussed in this section.

\subsubsection{Wide Survey}

The Wide Survey of Roman will cover the largest area of 2,702 deg$^2$, but done only in the $H$ band out to $m_{\rm AB}=26.2$. This added to the $H$ band coverage in several other Roman bands will enable CIB determination from over 5,100 deg$^2$ total. The reconstructed CIB from the remaining known galaxies is shown with thin, dashed green lines in Fig. \ref{fig:Roman_fluctuation_spectra}. It is sufficiently deep, covering a very large area to enable probing the CIB at $H$ band from new populations down to interesting levels. At scales $\gtrsim100''$ the ISM is expected to dominate the CIB fluctuations from known galaxies. However, if the amplitude of the new populations follows a Rayleigh-Jeans spectrum to $\sim 1$ $\mu$m then they should be safely above any ISM contributions. Furthermore, given the extensive sky coverage we will be able to isolate regions of the sky with low levels of ISM power which would decrease their relative contributions.

\subsubsection{Medium Survey}

The Medium Survey will cover 2,415 deg$^2$ in the Roman $Y$, $J$, $H$ bands to $m_{\rm AB}=26.5$ ($Y$) and 26.4 ($J$, $H$). The reconstructed source-subtracted CIB fluctuations from remaining known galaxies are shown with thin, solid green lines in Fig. \ref{fig:Roman_fluctuation_spectra}. The expected cirrus contributions are shown with shaded dash-dotted black lines and should be comfortably below the expected signal from the new populations at scales $\lesssim{10^4}''$ if their fluctuations scale according to the Rayleigh-Jeans energy spectrum. Additionally, the large sky area covered in the survey will allow us to select portions of the sky where ISM power is subdominant with respect to the signal from new populations. Furthermore, the residual CIB fluctuations from the remaining known galaxies are even lower than contributions from the ISM (at least on scales $\gtrsim100''$) which will enable us to constrain the signal from new populations to sub-percent statistical accuracy. This signal can then be analyzed via Lyman-tomography to robustly probe the history of emissions covered by the $Y$, $J$, $H$ bands, or $10<z<20$ \citep[see Fig. 1 of][]{Kashlinsky2015}.

\subsubsection{Deep Survey}

The 19.2 deg$^2$ covered here in the Roman $z$, $Y$, $J$, $H$, $F$, $K$ bands to $m_{\rm AB}=27.7$ (in $z,\,Y$), 27.6 ($J$), 27.5 ($H$), 27.0 ($F$ centered at 1.8 $\mu$m), 25.9 ($K$) would enable us to probe the CIB power and energy spectra over a wide range of scales and wavelengths down to the limits imposed by the reconstructed contributions from the remaining known galaxies shown with thick solid lines in the corresponding panels of Fig. \ref{fig:Roman_fluctuation_spectra}. The ISM contributions are shown with gray lines and may become significant at scales $\gtrsim 0.5^\circ$ at the shortest wavelengths. Importantly, the medium survey will cover 6 near-IR bands from 0.9 to 2.2 $\mu$m and can be evaluated after eliminating known sources to very faint magnitudes shown in Table \ref{tab:forthcoming}.
 
\subsubsection{Ultra Deep Survey}

The Ultra Deep Survey covers 5 deg$^2$ in Roman $Y$, $J$, $H$ bands to $m_{\rm AB}=28.2$ ($Y,\, J$) and 28.1 ($H$), making it the smallest of the surveys with respect to area. Shown in the corresponding panels of Fig. \ref{fig:Roman_fluctuation_spectra} are the contributions from the remaining known galaxies. The area covered should allow for a statistically meaningful determination of the source-subtracted CIB out to degree scales. Importantly, this Roman configuration should reach shot noise levels from known galaxy populations of only $P_{\rm SN} \simeq 1.8-1.1$ nJy$\cdot$nW m$^{-2}$ sr$^{-1}$ at the Roman $Y$, $J$, and $H$ bands and is unique for probing the dependence of the new populations' CIB power on the remaining shot noise, thereby probing new populations down to fluxes as faint as $S\sim P_{\rm SN}/F \sim (1-2)$ nJy \citep{Kashlinsky2007b}.

\subsubsection{High-Latitude Time Domain Survey}

The High-Latitude Time Domain Survey (HLTDS) covers 18.27 and 6.47 deg$^2$ corresponding to the Wide and Deep tiers respectively (\citealt{zasowski2025romanobservationstimeallocation}). The Wide tier includes observations in the Roman $R$, $z$, $Y$, $J$, $H$ bands to $m_{\rm AB}=28.8$ ($R$), 28.1 ($z$), 28.0 $Y$, 28.4 ($J$), and 28.9 ($H$) (Jeffrey Kruk, private communication). Imaging in the Deep tier on the other hand is done in the Roman $z$, $Y$, $J$, $H$, $F$ bands to $m_{\rm AB}= 29.4$ ($z$), 29.0 $Y,\,J$, 29.2 ($H$), and 29.3 ($F$). The expected contributions from faint galaxies below these magnitude limits are shown in the corresponding panels of Fig. \ref{fig:Roman_fluctuation_spectra}. As provided in Table \ref{tab:forthcoming}, these Roman configurations should reach shot noise levels from known galaxy populations of $P_{\rm SN} \lesssim 2.6$ nJy$\cdot$nW m$^{-2}$ sr$^{-1}$. The wavelength range, covered area(s), and sufficient depths can be used to probe contributions from new populations very efficiently via Lyman-tomography.

\subsection{SPHEREx}

\begin{figure*}[ht]
    \centering
     \includegraphics[width=\textwidth]{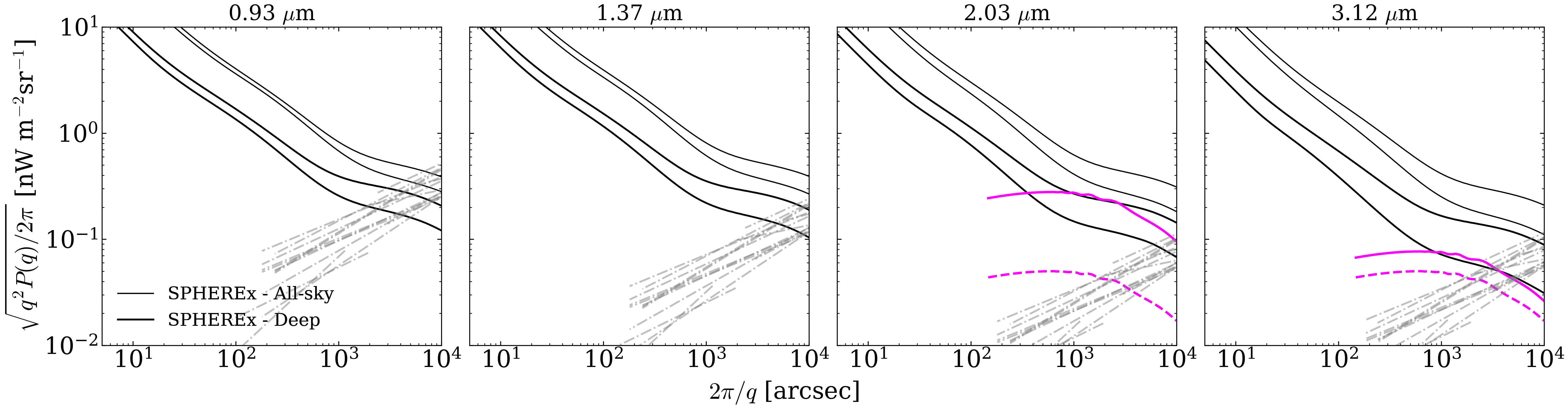}
   % \caption{CIB from known galaxies populations remaining in SPHEREx data is shown with black solid lines. The all-sky survey case is shown with black thin lines and the deep survey with thick solid lines. Upper lines correspond to the HFE limit and the lower ones to the LFE limit. At the 3.12 $\mu$m panel the pink lines show the source-subtracted CIB fluctuations from { Spizer} measurements \cite{Kashlinsky2025} at 3.6 $\mu$m (dashed line) and assuming that is scales $\propto \lambda^{-3}$ toward the central wavelength. The CIB from the remaining known populations overwhelms the signal from the new sources detected in Spitzer data.}
  \caption{Reconstructed SPHEREx-observed fluctuation spectra due to known galaxy populations are shown at 0.93, 1.37, and 2.03, and 3.12 $\mu$m. The All-sky Survey is shown with the solid, thin black lines while the Deep Survey is shown in the solid, thick black lines. The dash-dotted gray lines are estimated contributions from the ISM (see Sec. 3 of \cite{Kashlinsky2019}). The large-scale power seen in the 2.03 and 3.12 $\mu$m plots shows the estimates of the CIB signal from the Spitzer/Akari-found new populations assuming, for illustration, their high-$z$ origin; the amplitude marks the limits needed to probe the sought-for signal  where (1) the solid pink line assumes a $\Lambda$CDM model with an amplitude that scales with a Rayleigh-Jeans spectrum (normalized to the amplitude of the Spitzer 3.6 $\mu$m fluctuations) and (2) the dashed pink line shows the same $\Lambda$CDM model with an amplitude equal to that of the Spitzer 3.6 $\mu$m fluctuations.}
    \label{fig:SPHEREx_fluctuation_spectra}
\end{figure*}
SPHEREx \citep[the Spectro-Photometer for the History of the Universe, Epoch of Reionization, and ices Explorer,][]{Dore2014,Bock2025} is a NASA mission launched Mar 2025. Its spectrophotometer has 102 spectral channels distributed in 6 bands (0.73-1.13, 1.08-1.67, 1.61-2.45, 2.38-3.87, 3.79-4.44, 4.39-5.02 $\mu$m), yielding a spectral resolving power $\sim 40$ between 0.75 and 3.8 $\mu$m and $\sim 120$ between 3.8 and 5 $\mu$m. 
Spatially, SPHEREx samples the sky with $6.2''$ pixels, with 
each band spanning a $3.5^\circ \times 3.5^\circ$ field of view. 
Using standard 112.5 second exposures \citep{SPHEREX:2024},
SPHEREx can survey the full sky every 6 months, and a total
of 4 times during its planned 2-year mission.
At $\lambda<3.8$ $\mu$m, the combined all-sky survey will reach the sensitivity of $m_{\rm AB}\simeq 18.2-19.7$, and for an area of $\sim 200$ deg$^2$ surrounding the Ecliptic poles the SPHEREx deep survey will reach a sensitivity of $m_{\rm AB}\simeq 21-21.5$ \citep{SPHEREX:2024}. These expected (pre-launch) limits per spectral channel for single exposures, scaled by a factor of 2 to match planned, full survey depths are given in Table \ref{tab:forthcoming} \citep{Crill:2025}. The remaining CIB flux and, especially, shot noise levels from known galaxy populations are very much larger than what is needed in probing the origin of the CIB from new populations uncovered in Spitzer-based measurements. 
At wavelengths $\lambda>3.8$ $\mu$m the sensitivities are over a magnitude lower than in the 4 bands covering the shorter wavelengths and are too low for the study of the source-subtracted CIB envisaged here, so we do not present the reconstructed CIB from known populations there.

% {\color{red} If you average over {\bf bands} then you combine
% 17 {\bf channels} with a nominal gain of $>4$ in sensitivity  (1.5 mag) 
% compared to the single channel limits cited in Table 3.
% Was figure 14 constructed using single-channel sensitivities?}
For four spectral channels at wavelengths 
of $0.93,\, 1.37,\, 2.03,$ and $3.12$ $\mu$m, we show in Fig. \ref{fig:SPHEREx_fluctuation_spectra} the reconstructed CIB fluctuations from known galaxy populations at the magnitude limits provided in Table \ref{tab:forthcoming}. The Spitzer-detected source-subtracted anisotropies at 3.6 $\mu$m are shown with the pink dashed line at 3.12 $\mu$m; the pink solid line there shows the source-subtracted CIB component assuming a Rayleigh-Jeans energy spectrum. As seen in Fig. \ref{fig:SPHEREx_fluctuation_spectra}, the CIB power from the remaining known populations in SPHEREx exposures overwhelms the source-subtracted signal identified in the deep Spitzer maps.  The same limitation applies to using the Lyman-tomography methodology originally proposed for Euclid by \cite{Kashlinsky2015}. One would need to go about $\gtrsim 2$ magnitudes deeper reaching close to the Euclid Wide Survey magnitude limits.

\section{Summary}
In this work we use the newest and earlier CIB fluctuation maps from Spitzer to advance the empirical reconstruction of \cite{Helgason2012}. Doing so allows us to characterize the clustering of galaxies unresolved by Spitzer over a broad range of magnitude depths. This significantly more precise reconstruction, has important consequences for understanding the clustering and evolution of known galaxies. It also has implications regarding surveys with new and upcoming space missions such as Euclid, Roman, and SPHEREx. We summarize our findings below:
\begin{enumerate}
    \item We show that known galaxy populations are responsible for the source-subtracted CIB fluctuations at $\lesssim 1'$. These galaxies are hosted in DM halos of $\log(M_{\min}/M_{\odot}h^{-1})\gtrsim 11$, with this threshold increasing with shot noise $P_{\rm SN}$. These results are consistent with previous measurements of the unresolved CIB anisotropies \citep{Helgason2017,Kaminsky2025}. We use the flexibility of our machinery to implement different assumptions in our halo modeling and find that our results are largely insensitive to these changes relative to the uncertainties in the measurements and the width of the LFE and HFE bounds. Directly narrowing the LFE/HFE limits will be the focus of a follow-up investigation.
    \item Our results imply that the source-subtracted CIB fluctuations detected in the suite of Spitzer data at 3.6 and 4.5 $\mu$m over a range of angular scales between $30''\lesssim2\pi/q<(1-2)^\circ$ arise from necessarily faint sources contributing negligibly to the shot noise levels remaining in the data \citep{Kashlinsky2007b}. The large-scale fluctuations must then arise from faint, new cosmological objects contributing little to the observed small-scale structure, which is the sum of shot noise and 1-halo power, as required by their high-$z$ origin.

    \item The reconstructed angular power spectra due to known galaxies produce most of the small-scale power measured by Spitzer at 3.6 and 4.5 $\mu$m which disfavors any significant contributions from IHL proposed in \cite{Cooray2012} (see the IHL models shown in Fig. 22 of \cite{Kashlinsky2018} and references therein). The IHL model in general is also in conflict with the observed CXB-CIB coherence, first uncovered by \cite{Cappelluti2013}, which requires a significant abundance of accreting black holes among the new sources producing the large-scale source-subtracted CIB fluctuations.
    \item Using our Spitzer-based reconstructions we evaluate the angular power spectrum of known galaxy populations in sky surveys of the new and upcoming space missions, such as Euclid, Roman, and SPHEREx to higher accuracy than before. Of these missions, Roman has the best prospects of probing the nature of the new populations behind the Spitzer source-subtracted CIB signal, followed by Euclid, whereas in the SPHEREx surveys the known galaxies produce CIB fluctuations of substantially larger amplitude than expected from the new sources.
\end{enumerate}
\begin{acknowledgments}
We particularly thank Kari Helgason for very useful discussions over the years and for 
the current software for the LIBRAE project (\url{https://euclid.caltech.
edu/page/kashlinsky-team}) to reproduce CIB anisotropies
from known galaxy populations on which this development was based. A.J.K. acknowledges  Zhongtian Hu and Alberto Magaraggia for providing support on parallel computing which made the overall machinery much more efficient. A.J.K. also thanks Louise Paquereau and Ashley Ross for their input regarding the halo model formalism and assumptions. We thank Jeff Kruk for discussions of and information on the Roman surveys. This work was supported by NASA under award 80GSFC24M0006 and the authors acknowledge NASA/12-EUCLID11-0003 ``LIBRAE: Looking at Infrared Background Radiation Anisotropies with Euclid". A.K. and R.G.A. acknowledge
support from NASA award 80NSSC22K0621 ``Precision
measurement of source-subtracted cosmic infrared background
from new Spitzer data". A.J.K. acknowledges support from the NASA FINESST Program, Grant 80NSSC25K0309. A.J.K. and N.C. acknowledge the University of Miami for partial support.
\end{acknowledgments}

\software{astropy \citep{2013A&A...558A..33A,2018AJ....156..123A,2022ApJ...935..167A},  
          Colossus \citep{Diemer2018}, 
          halomod \citep{2021A&C....3600487M},
          hmf \citep{2013A&C.....3...23M}
          }
          
\bibliography{sample701,comments_bib}{}
\bibliographystyle{aasjournal}

\appendix
\section{Calibration systematics}\label{sec:cal_syst}

Here we characterize the level of residual calibration errors in the data, and 
demonstrate that they should not be a problem for the present analysis.
\cite{O'Brien:2025} have demonstrated a sensitive means for revealing flat field
errors in HST data that were used for analysis their study of the zodiacal 
light foreground. This involved mapping out the darkest regions across the 
detector for sets of thousands of images in a given filter. The results 
revealed patterns similar to the flat field, and the impact was quantified by 
examining the difference between the brightest and darkest regions across these
source-subtracted images.

\begin{figure*}[t] 
   \centering
   \includegraphics[width=7in]{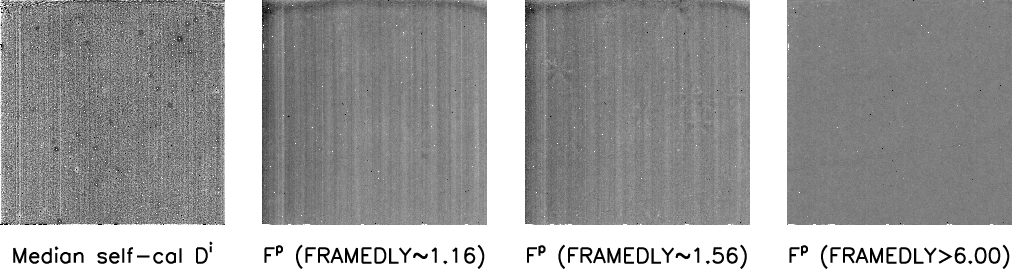} \\
   ~~\\
   \includegraphics[width=7in]{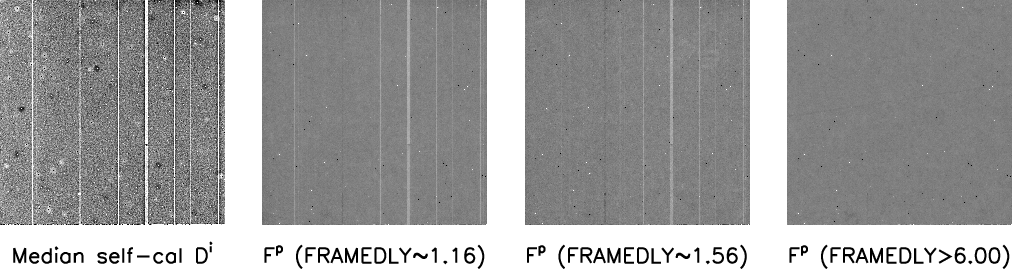} \\
   ~~\\
   \includegraphics[width=7in]{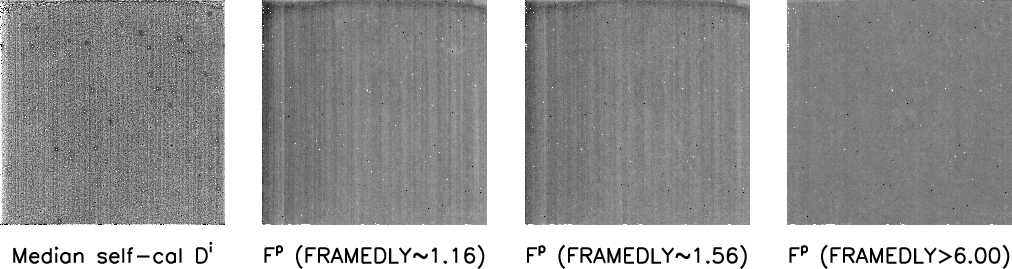} \\
   ~~\\
   \includegraphics[width=7in]{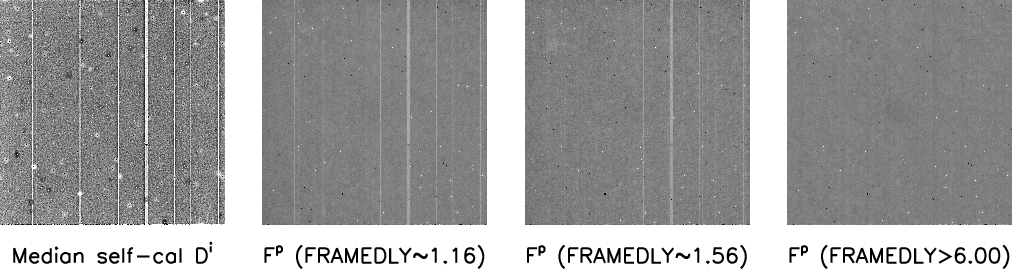} 
   \caption{The medians over all self-calibrated exposures ($D^i$) for a given field and wavelength 
   are shown in the left column ([-0.002,0.002] MJy sr$^{-1}$). Rows show
   results for NEP 3.6 $\mu$m, NEP 4.5 $\mu$m, CDFS 3.6 $\mu$m, and CDFS 4.5 $\mu$m. The other three columns 
   show examples of the derived self-calibration offsets ($F^p$) that are 
   applied to data with shorter or longer delay times (in s) for successive frames ([-0.1,0.1] MJy sr$^{-1}$).
   The residuals do resemble the self-calibration offsets for shorter 
   frame delay times, but are $\sim50$ times small in contrast.
   \label{fig:comment1}}
\end{figure*}

We applied the same technique to the IRAC LIBRAS data used here, and also find a
residual imprint of the instrument. However for quantifying the errors, rather
than 2-D histograms of the darkest subregions, we show the median sky brightness
across the detector for the full sets of $\sim 20000$ images in each band and each
field (Fig. \ref{fig:comment1}). 
These medians have a similar appearance to the histograms, but are 
useful in units of physical surface brightness, rather than expressing a 
normalized fraction. Also shown in Fig. \ref{fig:comment1}, are
examples of the self-calibrated detector offset, $F^p$, for the three 
groupings of frame delay times (time gaps between successive exposures, shown in the panels as FRAMEDLY in sec) that 
were used in the self-calibration \citep{Kashlinsky:2025}. To make the images 
comparable, the display range for the median images is 50 times tighter than for 
the offsets. It is evident that the residual medians are correlated with the 
self-calibration detector offsets applied for the short delay time data 
(half of each data set data), but not with those applied to long frame delay times.  
 
There is very little contrast in the background of the data being self-calibrated, 
and thus to avoid degeneracy between detector gain ($G^p$) and offset ($F^p$)
terms the self-calibration only solved for the latter \citep{Arendt:2010}.
We note that the IRAC flat fields had been found to be very 
stable, and that the superflats used for the pipeline calibration are estimated
to be accurate (for the warm-mission era) to 0.17\%, 0.09\% (1$\sigma$) at 3.6, 4.5 $\mu$m
\cite[Section 4.2 of][]{IRAC:2021}. 

For both the NEP and the CDFS, we find that the 1$\sigma$ dispersions of the median images are 
$\sim0.36$ and $\sim0.45$ nW~m$^{-2}$~sr$^{-1}$ at 3.6 and 4.5 $\mu$m. These are $<4\%$ of 
the mean noise level in the individual exposures. 
Thus, given that the resulting source-subtracted images have $\sim60$ exposures per
location, including highly varied dithers and large changes in position angle, 
errors in the detector calibration should be averaged down by a factor of 
$\sim \sqrt{60}$. At this level, the residual errors are not likely to 
contribute to either shot noise or 
large scale power in the analysis of the LIBRAS fields.

\end{document}